\documentclass[11pt]{article}
\usepackage[utf8]{inputenc}
\usepackage{amsfonts}
\usepackage{amsmath}
\usepackage{amssymb}
\usepackage{indentfirst}
\usepackage{graphicx}
\usepackage[dvipsnames]{xcolor}
\usepackage[colorlinks]{hyperref}
\usepackage{cite}
\usepackage{array}
\usepackage{microtype}
\usepackage{soul}
\usepackage[normalem]{ulem}
\usepackage{cancel}
\usepackage{soul}
\usepackage{xspace}
\usepackage{colortbl}
\usepackage{caption}
\newcolumntype{C}[1]{>{\centering\arraybackslash}p{#1}}
\definecolor{maroon}{cmyk}{0,0.87,0.68,0.32}

\numberwithin{equation}{section}
\allowdisplaybreaks

\makeatletter

\begin{document}

\begin{titlepage}
\vspace{3cm}

\baselineskip=24pt

\begin{center}
\textbf{\LARGE Extended Kinematical 3D Gravity Theories}
\par\end{center}{\LARGE \par}

\begin{center}
	\vspace{1cm}
	\textbf{Patrick Concha}$^{\ast, \bullet}$,
        \textbf{Daniel Pino}$^{\star, \bullet}$,
        \textbf{Lucrezia Ravera}$^{\dag, \ddag, \bullet}$,
	\textbf{Evelyn Rodríguez}$^{\ast, \bullet}$,
	\small
	\\[5mm]
    $^{\ast}$\textit{Departamento de Matemática y Física Aplicadas, }\\
	\textit{ Universidad Católica de la Santísima Concepción, }\\
\textit{ Alonso de Ribera 2850, Concepción, Chile.}
\\[2mm]
$^{\bullet}$\textit{Grupo de Investigación en Física Teórica, GIFT, }\\
	\textit{Concepción, Chile.}
\\[2mm]
	$^{\star}$\textit{Departamento de Física, Universidad de Concepción,}\\
	\textit{Casilla 160-C, Concepción, Chile.}
\\[2mm]
	$^{\dag}$\textit{DISAT, Politecnico di Torino, }\\
	\textit{ Corso Duca degli Abruzzi 24, 10129 Torino, Italy.}
	\\[2mm]
	$^{\ddag}$\textit{INFN, Sezione di Torino, }\\
	\textit{ Via P. Giuria 1, 10125 Torino, Italy.}
	 \\[5mm]
	\footnotesize
	\texttt{patrick.concha@ucsc.cl},
        \texttt{dpino2018@udec.cl},
        \texttt{lucrezia.ravera@polito.it},
	\texttt{erodriguez@ucsc.cl},
	\par\end{center}
\vskip 26pt
\begin{abstract}
In this work, we classify all extended and generalized kinematical Lie algebras that can be obtained by expanding the $\mathfrak{so}\left(2,2\right)$ algebra. We show that the Lie algebra expansion method based on semigroups reproduces not only the original kinematical algebras but also a family of non- and ultra-relativistic algebras. Remarkably, the extended kinematical algebras obtained as sequential expansions of the AdS algebra are characterized by a non-degenerate bilinear invariant form, ensuring the construction of a well-defined Chern-Simons gravity action in three spacetime dimensions. Contrary to the contraction process, the degeneracy of the non-Lorentzian theories is avoided without extending the relativistic algebra but considering a bigger semigroup. Using the properties of the expansion procedure, we show that our construction also applies at the level of the Chern-Simons action.

\end{abstract}
\end{titlepage}\newpage {} 

{\baselineskip=12pt \tableofcontents{}}

\section{Introduction}

The kinematical or space-time symmetry algebras play an important role in the construction of physical theories. Based on reasonable general assumptions, Bacry and Lévy-Leblond (1968) presented a classification of all kinematical algebras that establish connections between distinct inertial frames of reference \cite{Bacry:1968zf}. The Lie kinematical generators are interpreted as time translations, space translations, rotations and boosts. The kinematical Lie algebras contain, in addition to the relativistic AdS and Poincaré algebras, non-Lorentzian algebras which have received a renewed interest due to their diverse physical applications. In particular, non-relativistic algebras manifest within holography \cite{Son:2008ye,Balasubramanian:2008dm,Kachru:2008yh,Taylor:2008tg,Bagchi:2009my,Hartnoll:2009sz,Bagchi:2009pe,Christensen:2013lma,Christensen:2013rfa,Hartong:2014oma,Hartong:2014pma,Zaanen:2015oix}, Ho\v{r}ava-Lifshitz gravity \cite{Horava:2009uw,Hartong:2015wxa,Hartong:2015zia,Taylor:2015glc,Hartong:2016yrf,Devecioglu:2018apj}, effective field theory description of the quantum Hall effect \cite{Hoyos:2011ez,Son:2013rqa,Abanov:2014ula,Geracie:2014nka,Gromov:2015fda}, among others. Conversely, the ultra-relativistic symmetries have garnered contemporary attention due to their applications in tachyon condensation \cite{Gibbons:2002tv}, warped conformal field theories \cite{Hofman:2014loa}, tensionless strings \cite{Bagchi:2013bga, Bagchi:2015nca, Bagchi:2016yyf, Bagchi:2017cte, Bagchi:2018wsn}, holography in asymptotically flat space-times \cite{Barnich:2010eb,Barnich:2012aw,Bagchi:2010zz,Hartong:2015xda,Hartong:2015usd,Bagchi:2016bcd,Donnay:2022aba,Saha:2022gjw,Saha:2023hsl,Saha:2023abr}, asymptotic symmetries \cite{Perez:2021abf,Perez:2022jpr,Fuentealba:2022gdx} and in the context of black hole solutions \cite{Hawking:2016msc,Hawking:2016sgy,Donnay:2019jiz,Ciambelli:2019lap,Grumiller:2019fmp,deBoer:2023fnj,Ecker:2023uwm}.

On the other hand, the three-dimensional Chern-Simons (CS) gravity formalism provides a fertile testing ground for studying different aspects of higher-dimensional gravitational models as well their black hole solutions along with their thermodynamics \cite{Banados:1992wn}. In particular, Einstein-Hilbert action with or without cosmological constant can be written as a CS action considering the kinematical Lie algebras $\mathfrak{so}\left(2,2\right)$ or the $\mathfrak{iso}\left(2,1\right)$ \cite{Achucarro:1987vz,Witten:1988hc,Zanelli:2005sa}. The asymptotic symmetries of such models, appearing after imposing suitable boundary conditions \cite{Brown:1986nw}, result to be described by the conformal or the $\mathfrak{bms}_3$ algebra \cite{Ashtekar:1996cd,Barnich:2006av,Duval:2014uva}. The asymptotic analysis of three-dimensional CS gravity theories showcases one of the most highly investigated examples of the well-celebrated AdS/CFT duality \cite{Maldacena:1997re,Gubser:1998bc,Witten:1998qj}. 

One may then ask if a three-dimensional CS action can be constructed for all the kinematical Lie algebra of Bacry and Lévy-Leblond \cite{Bacry:1968zf}. An answer to such question was presented in \cite{Matulich:2019cdo} by considering that not all kinematical Lie algebras allow to obtain a well-defined CS action due to the degeneracy appearing in the non-relativistic regime. The question about which Lie algebra admits a non-degenerate invariant tensor is not new and diverse strategies have been adopted to overcome such difficulty. In the non-relativistic context, two central charges have to be added to the Newton-Hooke and Galilei algebra to construct a proper CS action in three spacetime dimensions. The new symmetries are known as extended Newton-Hooke \cite{Aldrovandi:1998im,Gibbons:2003rv,Brugues:2006yd,Alvarez:2007fw,Papageorgiou:2010ud,Duval:2011mi,Duval:2016tzi} and extended Bargmann \cite{Papageorgiou:2009zc,Bergshoeff:2016lwr}, the latter being a central extension of the Bargmann algebra. Such algebras can be alternatively be obtained as a contraction of the $\mathfrak{so}\left(2,2\right)\oplus\mathfrak{u}\left(1\right)^{2}$ and the $\mathfrak{iso}\left(2,1\right)\oplus\mathfrak{u}\left(1\right)^{2}$, respectively. Hence, the original cube of \cite{Bacry:1968zf} is extended to a tesseract starting from AdS with two trivial central charges whose non-relativistic limit admits now a non-degenerate bilinear form \cite{Matulich:2019cdo}.

Given the recent applications of the kinematical algebras and the advantages of the three-dimensional CS gravity formalism, there are at least two questions that one could explore. First, one may ask if it is possible to extend the cube of Bacry and Lévy-leblond to kinematical superalgebras analogously to the spin-3 generalization presented in \cite{Bergshoeff:2016soe}. Secondly, one may ask whether kinematical superalgebras can be used to construct three-dimensional CS supergravity actions. Although supersymmetric extensions of different kinematical algebras are known in the literature \cite{Andringa:2013mma, Bergshoeff:2015ija, Bergshoeff:2016lwr,Ozdemir:2019orp, deAzcarraga:2019mdn, Ozdemir:2019tby, Concha:2019mxx, Concha:2020tqx, Concha:2020eam,Concha:2021jos,Concha:2021llq,Ravera:2022buz,Bergshoeff:2022iyb}, it is not trivial to obtain them as sequential contraction of a relativistic AdS superalgebra. Moreover, the question whether they can be used to construct a well-defined CS supergravity is conditioned to a non-degenerate bilinear trace. On the other hand, the extension to supersymmetry of the tesseract presented in \cite{Matulich:2019cdo} cannot be trivially applied. 

In this work, motivated by our questions, we present an alternative approach to solve the degeneracy problem appearing in the original cube of \cite{Bacry:1968zf}. Here, we only focus on the bosonic case hoping that the method employed and the results will be useful to answer our two original questions in a future work. Given that the non-degeneracy requires to consider extended non-relativistic kinematical algebras, they cannot be obtained as contractions of the original $\mathfrak{so}\left(2,2\right)$ algebra. It is well known that the contraction process maintains the dimension of the Lie algebra. Then, an expansion procedure \cite{Hatsuda:2001pp,deAzcarraga:2002xi,Izaurieta:2006zz,deAzcarraga:2007et} is required to get higher-dimensional algebras. In particular, we shall use the so-called semigroup expansion ($S$-expansion) method \cite{Izaurieta:2006zz} given its recent applications in the non-Lorentzian regime \cite{Concha:2019lhn,Penafiel:2019czp,Gomis:2019nih,Bergshoeff:2020fiz,Concha:2022muu,Caroca:2022byi,Concha:2022you,Concha:2022jdc}. Hence, we extend the cube of \cite{Bacry:1968zf} to extended kinematical algebras which can be obtained as sequential expansions of the AdS algebra considering a particular semigroup and resonant conditions. Remarkably, unlike \cite{Matulich:2019cdo}, the method employed here does not require central extensions of the original relativistic algebras to avoid degeneracy in the non-relativistic counterpart. Such a feature would be useful for generalizing our results in the presence of supersymmetry.

Then, we generalize our result to new cubes with generalized kinematical algebras which contain, as particular sub-cases, the Newtonian symmetry algebras, post-Newtonian extensions along with their ultra-relativistic versions. Finally, we construct the three-dimensional CS gravity actions based on the extended kinematical algebras equipped with non-degenerate invariant bilinear form. To this end, the S-expansion procedure provides us with the expanded invariant tensor required to construct CS actions in terms of the original ones, which is an additional motivation for considering this method. In this direction, it is important to emphasize that the S-expansion procedure does not guarantee the non-degeneracy of the expanded invariant tensor. For an arbitrary semigroup, the non-degeneracy of the invariant trace for a given expanded (super)algebra requires to be checked case-by-case. Nonetheless, the semigroups considered here to obtain the corresponding non-relativistic regime of the kinematical algebras along with the corresponding resonant conditions allow us to elucidate which generalized kinematical algebras admit a non-degenerate invariant tensor. 

The paper is organized as follows. In section \ref{sec2} we present a brief review of the kinematical algebras introduced in \cite{Bacry:1968zf}. Sections \ref{sec3} and \ref{sec4} contain our main results. Section \ref{sec3} is devoted to the generalization of the kinematical algebras by considering the semigroup expansion method. We first present extended kinematical Lie algebra characterized by a non-degenerate invariant tensor. In section \ref{sec4} we present the construction of the CS gravity actions based on the extended kinematical algebras. We conclude our work with some comments about future developments in Section \ref{sec5}.

%%%%%%%%%%%%%%%%%%%%%%%%%%%%%%%%%%%%%%%%%%%%%%%%%%%%%%%%%%%%%%%%%%%%%%%%%%%%%%%%%%%%%%%%%%%%%%%%%%%%%%%

\section{Kinematical algebras}\label{sec2}

\begin{center}
 \begin{figure}[h!]
  \begin{center}
        \includegraphics[width=9.3cm, height=8.3cm]{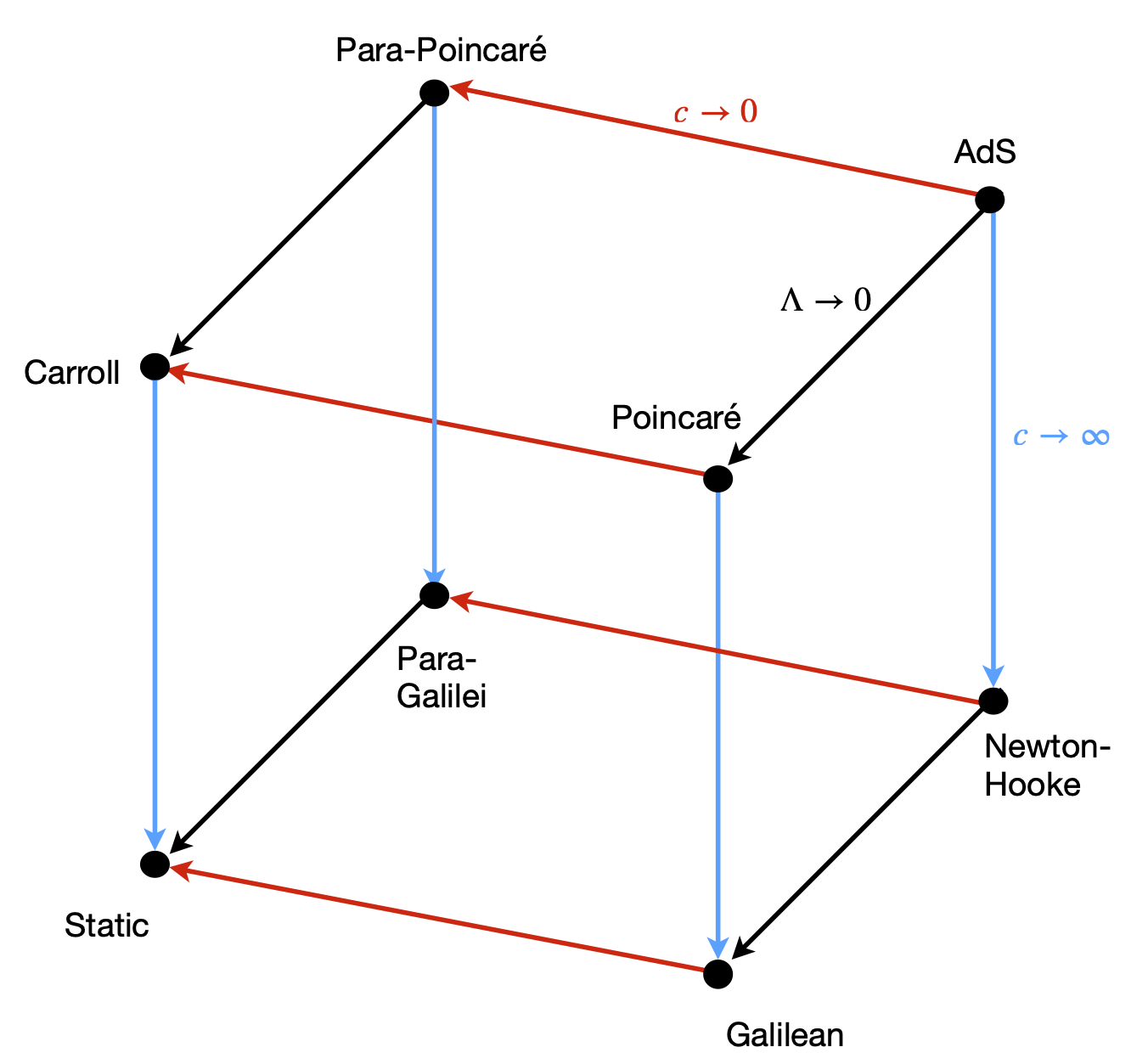}
        \captionsetup{font=footnotesize}
        \caption{This cube summarizes the different limits starting from the AdS Lie algebra.}
        \label{fig1}
         \end{center}
        \end{figure}
    \end{center}

In this section we briefly review the kinematical Lie algebras defined in three spacetime dimensions following the cube introduced by Bacry and Lévy-Leblond in \cite{Bacry:1968zf}. Starting from the $\mathfrak{so}\left(2,2\right)$ Lie algebra, one can derive different Lie algebras by applying successive Inönü-Wigner contractions which can be seen as particular limits leading to diverse physical regimes (see Figure \ref{fig1}). 

The well-known Poincaré algebra, given by the $\mathfrak{iso}\left(2,1\right)$ Lie algebra, appears as a vanishing cosmological constant limit $\ell\rightarrow 0$ where the AdS radius $\ell$ parameter is related to the cosmological constant through $\Lambda = - \frac{1}{\ell^2}$. Two inequivalent limits can be applied on the speed of light $c$. On one hand, the non-relativistic limit $c\rightarrow\infty$ of the relativistic AdS algebra defined in three spacetime dimensions reproduces the Newton-Hooke algebra, which in turn leads to the Galilean symmetry in the flat limit $\ell\rightarrow\infty$. On the other hand, the so-called para-Poincaré symmetry\footnote{Also called AdS-Carroll. Let us note that the de Sitter Carroll algebra, also denoted as the Lie algebra of the inhomogeneous $SO\left(4\right)$ group, is obtained when we consider the ultra-relativistic limit to $\mathfrak{so}\left(3,1\right)$ rather than $\mathfrak{so}\left(2,2\right)$.} appears as an ultra-relativistic limit $c\rightarrow 0$ of the $\mathfrak{so}\left(2,2\right)$ Lie algebra. In the vanishing cosmological constant limit $\ell\rightarrow\infty$, the ultra-relativistic limit $c\rightarrow 0$ of the Poincaré algebra reproduces the Carroll one. Interestingly, we obtain the so-called static algebra applying three successive contractions from the AdS Lie algebra. In summary, there are four different contraction processes starting from the $\mathfrak{so}\left(2,2\right)$ which have been denoted as space-time ($\ell\rightarrow\infty$), speed-space ($c\rightarrow\infty$), speed-time ($c\rightarrow 0$) and general contraction. In particular, the general contraction can be seen as a sequential of the other three contractions \cite{Bacry:1968zf}.

In order to visualize the diverse contractions, let us start with the $\mathfrak{so}\left(2,2\right)$ algebra whose commutation relations are given by
\begin{align}
\left[ \hat{J}_{A},\hat{J}_{B}\right] &=\epsilon _{ABC}\hat{J}^{C}\,, &  
\left[ \hat{J}_{A},\hat{P}_{B}\right] &=\epsilon _{ABC}\hat{P}^{C}\,,  &
\left[ \hat{P}_{A},\hat{P}_{B}\right] &=\epsilon _{ABC}\hat{J}^{C}\,.  \label{AdS}
\end{align}%
Here, $\hat{J}_{A}$ corresponds to the Lorentz generators and $\hat{P}_{A}$ are the spacetime translations. Let us note that the AdS algebra can be written, in three spacetime dimensions, as two copies of the $\mathfrak{so}\left(2,1\right)$ or $\mathfrak{sl}\left(2,\mathbb{R}\right)$ algebra. Before applying the diverse limits, it is convenient to decompose the relativistic Lorentz indices $A=0,1,2$ by considering a time-space splitting such that $A=\{0,a\}$ with $a=1,2$. Then, the AdS algebra is spanned by the set of generators $\{J,G_{a},H,P_{a}\}$ where we have relabeled the AdS generators as follow
\begin{align}
    J&=\hat{J}_{0}\,, & G_{a}&=\hat{J}_{a}\,, & H&=\hat{P}_{0}\,, & P_{a}&=\hat{P}_{a}\,. \label{split}
\end{align}
The rescaling of the relativistic AdS generators has to be considered as in Table \ref{Table1} in order to reproduce the different contractions. Then, the contracted algebra is obtained after performing the limit of the rescaling parameter going to infinity.

\begin{table}[h]
    \centering
    \begin{tabular}{|c||c|c|c|c|}
    \hline
      \rowcolor[gray]{0.9} Generators  & Space-time & Speed-space & Speed-time & General \\ \hline
        $J$ & $J$& $J$& $J$& $J$\\
       \rowcolor[gray]{0.9} $G_a$ & $G_a$& $c\,G_{a}$& $\tau\,G_{a}$& $c\tau\,G_{a}$\\
        $H$ & $\ell \,H$& $H$ & $\tau\,H$ & $\ell\tau\,H$ \\
       \rowcolor[gray]{0.9} $P_a$ & $\ell \,P_{a}$ & $c\, P_{a}$ &$P_{a}$ & $\ell c\, P_{a}$ \\
     \hline
         \end{tabular}
         \captionsetup{font=footnotesize}
    \caption{Rescaling of the generators for different contractions. Here $\tau=1/c$ reproduces the ultra-relativistic limit when $\tau\rightarrow\infty$.}
    \label{Table1}
\end{table}
\noindent The explicit commutation relations of the kinematical Lie algebras that can be obtained via sequential contraction procedures are listed in Tables \ref{Table2} and \ref{Table3}. We have intentionally separated our results in two tables in order to manifest which algebras are good algebras to construct three-dimensional CS actions. Indeed, the kinematical algebras of Table \ref{Table2} admit a non-degenerate bilinear invariant trace allowing to construct well-defined CS actions. The non-degeneracy of the action ensures a kinetic term for each gauge field and the field equations are given by the vanishing of the curvatures.

\begin{table}[h]
    \centering
    \begin{tabular}{|l||r|r|r|r|}
    \hline
      \rowcolor[gray]{0.9} Commutators  & AdS & Poincaré & Para-Poincaré & Carroll \\ \hline
        $\left[ J,G_{a} \right]$ & $\epsilon_{ab}G_{b}$ & $\epsilon_{ab}G_{b}$& $\epsilon_{ab}G_{b}$& $\epsilon_{ab}G_{b}$\\
        \rowcolor[gray]{0.9}$\left[ J,P_{a} \right]$ & $\epsilon_{ab}P_{b}$& $\epsilon_{ab}P_{b}$ & $\epsilon_{ab}P_{b}$ & $\epsilon_{ab}P_{b}$ \\
        $\left[ G_{a},G_{b} \right]$ & $-\epsilon_{ab}J$ & $-\epsilon_{ab}J$ &$0$ & $0$ \\
        \rowcolor[gray]{0.9}$\left[ H,G_{a} \right]$ & $\epsilon_{ab}P_{b}$ & $\epsilon_{ab}P_{b}$ &$0$ & $0$ \\
        $\left[ G_{a},P_{b} \right]$ & $-\epsilon_{ab}H$ & $-\epsilon_{ab}H$ &$-\epsilon_{ab}H$ & $-\epsilon_{ab}H$ \\
        \rowcolor[gray]{0.9}$\left[ H,P_{a} \right]$ & $\epsilon_{ab}G_{b}$ & $0$ &$\epsilon_{ab}G_{b}$ & $0$ \\
        $\left[ P_{a},P_{b} \right]$ & $-\epsilon_{ab}J$ & $0$ &$-\epsilon_{ab}J$ & $0$ \\
     \hline
         \end{tabular}
         \captionsetup{font=footnotesize}
    \caption{Commutation relations of the AdS, Poincaré, Para-Poincaré and Carroll algebra.}
    \label{Table2}
\end{table}

\begin{table}[h]
    \centering
    \begin{tabular}{|l||r|r|r|r|}
    \hline
     \rowcolor[gray]{0.9}  Commutators  & Newton-Hooke & Galilei & Para-Galilei & Static \\ \hline
        $\left[ J,G_{a} \right]$ & $\epsilon_{ab}G_{b}$ & $\epsilon_{ab}G_{b}$& $\epsilon_{ab}G_{b}$& $\epsilon_{ab}G_{b}$\\
       \rowcolor[gray]{0.9} $\left[ J,P_{a} \right]$ & $\epsilon_{ab}P_{b}$& $\epsilon_{ab}P_{b}$ & $\epsilon_{ab}P_{b}$ & $\epsilon_{ab}P_{b}$ \\
        $\left[ G_{a},G_{b} \right]$ & $0$ & $0$ &$0$ & $0$ \\
        \rowcolor[gray]{0.9}$\left[ H,G_{a} \right]$ & $\epsilon_{ab}P_{b}$ & $\epsilon_{ab}P_{b}$ &$0$ & $0$ \\
        $\left[ G_{a},P_{b} \right]$ & $0$ & $0$ &$0$ & $0$ \\
       \rowcolor[gray]{0.9}$\left[ H,P_{a} \right]$ & $\epsilon_{ab}G_{b}$ & $0$ &$\epsilon_{ab}G_{b}$ & $0$ \\
        $\left[ P_{a},P_{b} \right]$ & $0$ & $0$ &$0$ & $0$ \\
     \hline
         \end{tabular}
         \captionsetup{font=footnotesize}
    \caption{Commutation relations of the Newton-Hooke, Galilei, Para-Galilei and Static algebra.}
    \label{Table3}
\end{table}
 On the other hand, the kinematical algebras listed in Table \ref{Table3}, which are obtained through a speed-space contraction, suffer from degeneracy. Different strategies have been implemented to avoid degeneracy. In the next section we will show how this problem can be overcome using the Lie algebra expansion method based on semigroups \cite{Izaurieta:2006zz}. Such procedure provides us not only with the desired non-relativistic symmetries but also with the non-vanishing components of a non-degenerate invariant tensor which are required to construct an action in the CS formalism.
%%%%%%%%%%%%%%%%%%%%%%%%%%%%%%%%%%%%%%%%%%%%%%%%%%%%%%%%%
%%%%%%%%%%%%%%%%%%%%%%%%%%%%%%%%%%%%%%%%%%%%%%%%%%%%%%%%%

\section{Generalized kinematical algebras and S-expansion method}\label{sec3}

It is well known that, in three spacetime dimensions, the addition of two central charges in the non-relativistic Newton-Hooke and Galilei algebras allows for a non-degenerate invariant tensor. Here, we first present the minimal setup required to extend the cube to  kinematical Lie algebras admitting a non-degenerate invariant metric. We show that the space-time and speed-time contractions can be seen as particular $S$-expansions, in the sense that the expanded algebras have the same dimension of the original one. In other words, these particular expansions reproduce Inönü-Wigner contractions. The same holds for the speed-space case.  However, in order to obtain algebras endowed with a non-degenerate invariant tensor, the speed-space contraction has to be replaced by a different expansion yielding higher-dimensional Lie algebras. This is achieved by considering a larger semigroup $S$. We then present generalized kinematical algebras as sequential $S$-expansions starting from the $\mathfrak{so}\left(2,2\right)$ algebra. 

The $S$-expansion method basically consists in obtaining a new Lie algebra $\mathfrak{G}$ from a given one $\mathfrak{g}$ by combining the generators and structure constants of the original Lie algebra with the elements of a semigroup $S$. The expanded Lie algebra is then related to the original one as $\mathfrak{G}=S\times\mathfrak{g}$. A smaller subalgebra can be extracted from the $S$-expanded one by considering a resonant expansion or a $0_{S}$-reduction \cite{Izaurieta:2006zz}. Here, we shall show that considering a particular family of semigroups we are able not only to reproduce the cube of Bacry and Levy-Leblond \cite{Bacry:1968zf} but also to extend it to generalized kinematical algebras admitting a non-degenerate invariant bilinear trace.

\subsection{Kinematical Lie algebras}
A particular semigroup $S_{E}^{\left(1\right)}$ reproduces the Inönü-Wigner contraction. In particular, we shall see that the space-time, speed-space, speed-time and even the general contraction can be substituted by a $S_{E}^{\left(1\right)}$-expansion but considering different subspace decompositions of the original algebra. Before applying the S-expansion, let us consider first the $\mathfrak{so}\left(2,2\right)$ Lie algebra \eqref{AdS} with the relabeled generators as in \eqref{split}. Let $V_0$ and $V_1$ be two subspaces of the AdS algebra which, depending on the desired contraction, are given by Table \ref{Table4}.
\begin{table}[h]
    \centering
    \begin{tabular}{|c||c|c|c|c|}
    \hline
     \rowcolor[gray]{0.9}  Subspaces  & Space-time & Speed-space & Speed-time & General \\ \hline
        $V_0$ & $J,G_a$& $J,H$& $J,P_{a}$& $J$\\
       \rowcolor[gray]{0.9} $V_1$ & $H,P_a$& $G_{a},P_{a}$& $H,G_{a}$& $H,G_{a},P_{a}$\\
     \hline
         \end{tabular}
         \captionsetup{font=footnotesize}
    \caption{Subspaces decomposition of the AdS algebra.}
    \label{Table4}
\end{table}

\noindent One can check that the subspaces $V_0$ and $V_1$ satify a $\mathbb{Z}_2$-graded Lie algebra,
\begin{align}
[V_0,V_0]&\subset V_0\,,  &[V_0,V_1]&\subset V_1\,,  &[V_1,V_1]&\subset V_0\,.\label{sd}
\end{align}
Let us consider now $S_{E}^{\left(1\right)}=\{\lambda_0,\lambda_1,\lambda_2\}$ as the relevant semigroup whose elements satisfy the following multiplication law
\begin{equation}
\begin{tabular}{l|llll}

$\lambda _{2}$ & $\lambda _{2}$ & $\lambda _{2}$ & $\lambda _{2}$  \\
$\lambda _{1}$ & $\lambda _{1}$ & $\lambda _{2}$ & $\lambda _{2}$  \\
$\lambda _{0}$ & $\lambda _{0}$ & $\lambda _{1}$ & $\lambda _{2}$  \\ \hline
& $\lambda _{0}$ & $\lambda _{1}$ & $\lambda _{2}$ 
\end{tabular}
\label{mlSE1}
\end{equation}
Here $\lambda_2=0_S$ is the zero element of the semigroup which satisfies $0_S\lambda_i=\lambda_i 0_S=0_S$. Then, let $S_{E}^{\left(1\right)}=S_0\cup S_1$ be a subset decomposition with
\begin{align}
    S_0&=\{\lambda_0,\lambda_2\}\,, & S_1&=\{\lambda_1,\lambda_2\}\,, 
\end{align}
which is said to be resonant since it satifies the same algebraic structure than the subspaces decomposition of Table \ref{Table4}, namely
\begin{align}
S_0\cdot S_0&\subset S_0\,, \quad &S_0\cdot S_1&\subset S_1\,, \quad &S_1\cdot S_1&\subset S_0\,.\label{rc}
\end{align}
Then, each type of contraction can be recovered as a $S_{E}^{\left(1\right)}$-resonant expansion of the relativistic $\mathfrak{so}\left(2,2\right)$ algebra,
\begin{align}
    \mathfrak{G}=\left(S_0\times V_0\right)\oplus\left(S_1\times V_1\right)\,,
\end{align}
followed by a $0_S$-reduction, namely $0_S J=0_SG_a=0_SH=0_SP_a=0$.  The expanded generators are related to the AdS one through the semigroup elements as in Table \ref{Table5}.
\begin{table}[h!]
\renewcommand{\arraystretch}{1.3}
\centering
    \begin{tabular}{l ||C{1.5cm}|C{1.5cm}||C{1.5cm}|C{1.5cm}||C{1.5cm}|C{1.5cm}||C{0.9cm}|C{1.9cm}|}
    & \multicolumn{2}{|c||}{Space-time}& \multicolumn{2}{|c||}{Speed-space}& \multicolumn{2}{|c||}{Speed-time}& \multicolumn{2}{|c|}{General} \\ \hline
    $\lambda_2$ & \cellcolor[gray]{0.8}  & \cellcolor[gray]{0.8} & \cellcolor[gray]{0.8} & \cellcolor[gray]{0.8}& \cellcolor[gray]{0.8} &\cellcolor[gray]{0.8} & \cellcolor[gray]{0.8} & \cellcolor[gray]{0.8} \\ \hline 
    $\lambda_1$ & \cellcolor[gray]{0.8} & $\texttt{H}$, \, $\texttt{P}_a$ & \cellcolor[gray]{0.8} & $\texttt{G}_a$, \,$\texttt{P}_a$ & \cellcolor[gray]{0.8} & $\texttt{H}$, $\texttt{G}_a$ & \cellcolor[gray]{0.8} & $\texttt{H}$,\, $\texttt{G}_a$, \,$\texttt{P}_a$ \\ \hline
    $\lambda_0$  & $\texttt{J}$,\, $\texttt{G}_a$ & \cellcolor[gray]{0.8} & $\texttt{J}$,\, $\texttt{H}$ & \cellcolor[gray]{0.8} & $\texttt{J}$, $\texttt{P}_a$ & \cellcolor[gray]{0.8} & $\texttt{J}$ & \cellcolor[gray]{0.8} \\ \hline
    &$J$, $G_a$ & $H$, $P_a$ &$J$, $H$ & $G_a$, $P_a$ &$J$, $P_a$ & $H$, $G_a$ &$J$ & $H$, $G_a$, $P_a$ \\ 
    \end{tabular}
    \captionsetup{font=footnotesize}
    \caption{Expanded generators in terms of the AdS ones and the semigroup elements. The AdS generators have been organized according to the subspaces decomposition of Table \ref{Table4}.}
    \label{Table5}
    \end{table}

\begin{center}
 \begin{figure}[h!]
  \begin{center}
        \includegraphics[width=8.8cm, height=8.8cm]{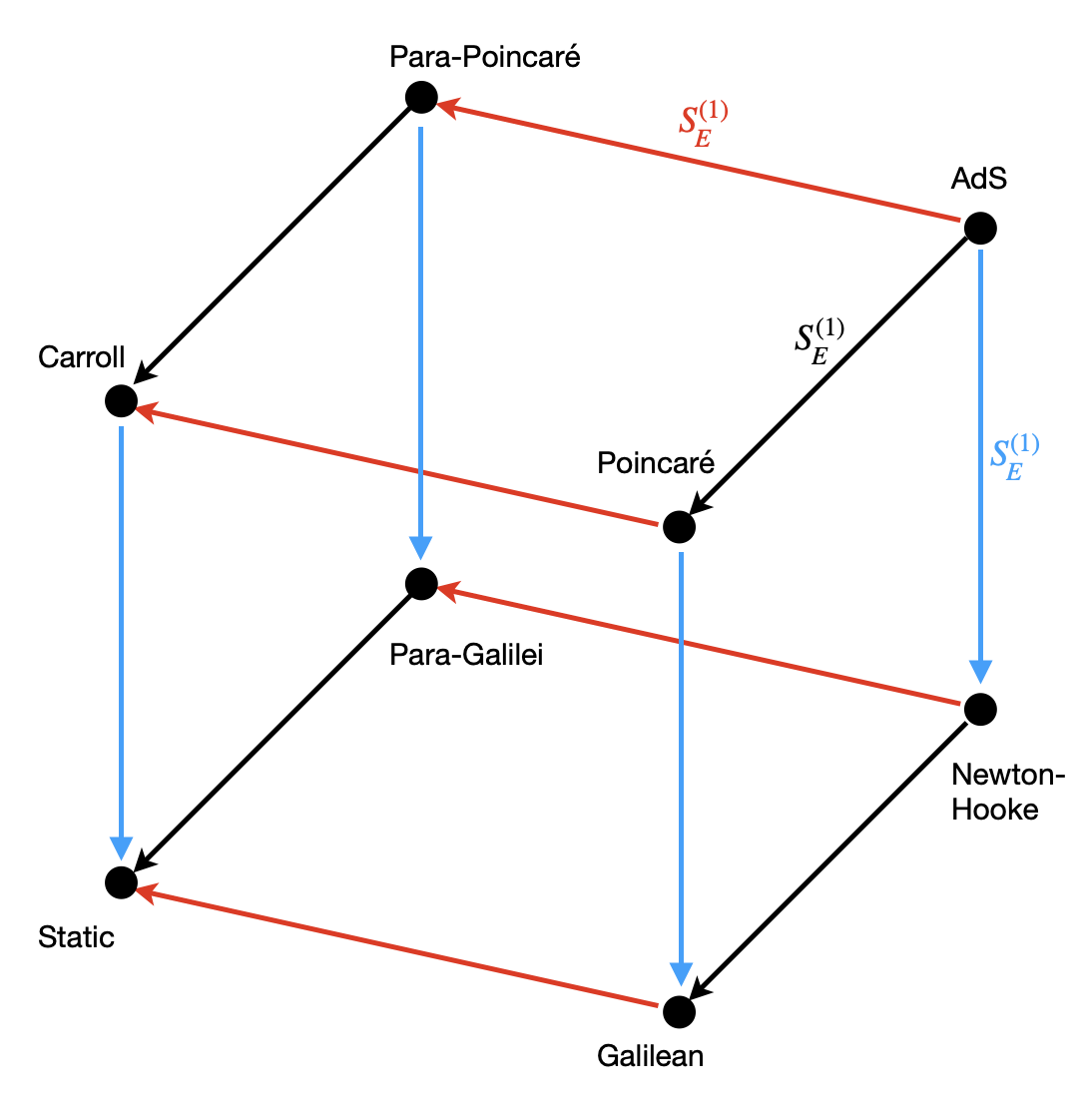}
        \captionsetup{font=footnotesize}
        \caption{This cube reproduces the original kinematical algebras as sequential expansions starting from the AdS algebra. Although the $S_{E}^{\left(1\right)}$ semigroup is applied in all the directions, there are three different subspace decompositions of the original algebra according to Table \ref{Table4}. Then, each expanded kinematical generator is related to an original one as in Table \ref{Table5}.}
        \label{figN}
         \end{center}
        \end{figure}
    \end{center}

The commutation relations for the expanded generators are obtained by considering the commutation relations of the AdS algebra and the multiplication law of the semigroup $S_{E}^{\left(1\right)}$. Then, one can easily check that the expanded generators $\{\texttt{J},\texttt{G}_{a},\texttt{H},\texttt{P}_a\}$ coming from the space-time subspaces decomposition satisfy the Poincaré algebra. On the other hand, the expanded algebra for the speed-space and speed-time cases are the Para-Poincaré and the Newton-Hooke algebras, respectively. Interestingly, the procedure can be applied starting from any kinematical algebra. For instance, one can obtain the Galilei and Carroll symmetry by applying the $S_{E}^{\left(1\right)}$ resonant expansion of the Poincaré algebra and performing a $0_S$-reduction. In such case the subspaces decomposition of the Poincaré algebra for recovering the non- and ultra-relativistic counterpart are analogue to those considered in Table \ref{Table4}, whose generators satisfy now the Poincaré commutators. Alternatively, the Carroll algebra can be obtained by $S_{E}^{\left(1\right)}$-expanding the Para-Poincaré algebra considering the resonant condition and after applying the $0_S$-reduction. However, in such case, the subspaces decomposition of the Para-Poincaré algebra is the one considered for reproducing a speed-space limit. Thus, the $S_{E}^{\left(1\right)}$ semigroup along the resonant condition and the $0_S$-reduction reproduce the sequential limits allowing to obtain the kinematical algebras appearing in the original cube (see Figure \ref{figN}). 

\subsection{Extended kinematical Lie algebras}
The previous procedure can be extended with a larger semigroup $S_{E}^{\left(2\right)}$ to avoid degeneracy in the non-relativistic regime. To this end, let us first consider the speed-space subspaces decomposition of the AdS algebra given by $V_0=\{J,H\}$ and $V_1=\{G_a,P_a\}$. Then, let $S_{E}^{\left(2\right)}=\{\lambda_0,\lambda_1,\lambda_2,\lambda_3\}$ be the relevant semigroup whose elements satisfy the following multiplication law,
\begin{equation}
\begin{tabular}{l|llll}
$\lambda _{3}$ & $\lambda _{3}$ & $\lambda _{3}$ & $\lambda _{3}$ & $\lambda
_{3}$ \\
$\lambda _{2}$ & $\lambda _{2}$ & $\lambda _{3}$ & $\lambda _{3}$ & $\lambda
_{3}$ \\
$\lambda _{1}$ & $\lambda _{1}$ & $\lambda _{2}$ & $\lambda _{3}$ & $\lambda
_{3}$ \\
$\lambda _{0}$ & $\lambda _{0}$ & $\lambda _{1}$ & $\lambda _{2}$ & $\lambda
_{3}$ \\ \hline
& $\lambda _{0}$ & $\lambda _{1}$ & $\lambda _{2}$ & $\lambda _{3}$%
\end{tabular}
\label{mlSE2}
\end{equation}

\noindent with $\lambda_3$ being the zero element of the semigroup. A resonant subset decomposition of the semigroup $S_{E}^{\left(2\right)}$ is given by $S_{E}^{\left(2\right)}=S_0\cup S_1$ with
\begin{align}
    S_0&=\{\lambda_0,\lambda_2,\lambda_3\}\,, & S_1&=\{\lambda_1,\lambda_3\}\,.\label{sd2}
\end{align}
Then a non-relativistic symmetry is obtained after applying the resonant $S_{E}^{\left(2\right)}$-expansion to the $\mathfrak{so}\left(2,2\right)$ algebra and performing its $0_S$-reduction. The non-relativistic generators are related to the relativistic AdS ones through the semigroup elements as in Table \ref{Table6}.
\renewcommand{\arraystretch}{0.8}
\begin{table}[h!]
\centering
    \begin{tabular}{l|C{1,6cm}|C{1,6cm}|}
%\cline{2-2}\cline{3-3}
& \multicolumn{2}{|c|}{Speed-space}\\ \hline
$\lambda_3$ & \multicolumn{1}{|l}{\cellcolor[gray]{0.8}} & \multicolumn{1}{|l|}{\cellcolor[gray]{0.8}} \\ \hline
$\lambda_2$ & $\texttt{S}$,\, $\texttt{M}$ & \multicolumn{1}{|l|}{\cellcolor[gray]{0.8}} \\ \hline
$\lambda_1$ & \multicolumn{1}{|l|}{\cellcolor[gray]{0.8}} & $\texttt{G}_a$,\ $\texttt{P}_a$ \\ \hline
$\lambda_0$ & $ \texttt{J}$,\, $\texttt{H}$ & \multicolumn{1}{|l|}{\cellcolor[gray]{0.8}} \\ \hline
 & $J$, $H$ & $G_a$, $P_a$
\end{tabular}
\captionsetup{font=footnotesize}
\caption{Non-relativistic generators in terms of the AdS ones and the semigroup elements.}
\label{Table6}%
\end{table}

\noindent The expanded commutation relations are obtained considering the multiplication law of the semigroup $S_{E}^{\left(2\right)}$ along the original AdS commutators \eqref{AdS}. In particular, the expanded generators satisfy the extended Newton-Hooke algebra \cite{Aldrovandi:1998im,Gibbons:2003rv,Brugues:2006yd,Alvarez:2007fw,Papageorgiou:2010ud,Duval:2011mi,Duval:2016tzi} whose commutation relations can be found in Table \ref{Table7}. Following the same procedure, the extended Bargmann algebra \cite{Papageorgiou:2009zc,Bergshoeff:2016lwr} can be obtained starting from the relativistic Poincaré algebra with the speed-space decomposition of Table \ref{Table6} and the $S_{E}^{\left(2\right)}$ semigroup. Moreover, extended versions of a para-Bargmann algebra and the static one are derived considering the resonant $S_{E}^{\left(2\right)}$-expansion of the Carroll and para-Poincaré algebra, respectively. Here, the para-Bargmann algebra denotes a central extension of the para-Galilei algebra which results to be isomorphic to the Bargmann algebra but physically different. The four obtained extended kinematical algebras appear considering a resonant $S_{E}^{\left(2\right)}$-expansion, a $0_S$-reduction and the speed-space decomposition of the kinematical algebra of the Table \ref{Table2}. Such extended kinematical algebras, as we shall see in section \ref{sec4}, admit a non-degenerate invariant bilinear trace due to the presence of two central charges $\texttt{S}$ and $\texttt{M}$. 
\begin{table}[h]
    \centering
    \begin{tabular}{|l||r|r|r|r|}
    \hline
      \rowcolor[gray]{0.9} Commutators  & Extended  & Extended  & Extended  & Extended \\ 
      \rowcolor[gray]{0.9} & Newton-Hooke & Bargamnn & para-Bargmann & static \\ \hline
        $\left[ \texttt{J},\texttt{G}_{a} \right]$ & $\epsilon_{ab}\texttt{G}_{b}$ & $\epsilon_{ab}\texttt{G}_{b}$& $\epsilon_{ab}\texttt{G}_{b}$& $\epsilon_{ab}\texttt{G}_{b}$\\
       \rowcolor[gray]{0.9} $\left[ \texttt{J},\texttt{P}_{a} \right]$ & $\epsilon_{ab}\texttt{P}_{b}$& $\epsilon_{ab}\texttt{P}_{b}$ & $\epsilon_{ab}\texttt{P}_{b}$ & $\epsilon_{ab}\texttt{P}_{b}$ \\
        $\left[ \texttt{G}_{a},\texttt{G}_{b} \right]$ & $-\epsilon_{ab}\texttt{S}$ & $-\epsilon_{ab}\texttt{S}$ &$0$ & $0$ \\
       \rowcolor[gray]{0.9} $\left[ \texttt{H},\texttt{G}_{a} \right]$ & $\epsilon_{ab}\texttt{P}_{b}$ & $\epsilon_{ab}\texttt{P}_{b}$ &$0$ & $0$ \\
       $\left[ \texttt{G}_{a},\texttt{P}_{b} \right]$ & $-\epsilon_{ab}\texttt{M}$ & $-\epsilon_{ab}\texttt{M}$ &$-\epsilon_{ab}\texttt{M}$ & $-\epsilon_{ab}\texttt{M}$ \\
       \rowcolor[gray]{0.9} $\left[ \texttt{H},\texttt{P}_{a} \right]$ & $\epsilon_{ab}\texttt{G}_{b}$ & $0$ &$\epsilon_{ab}\texttt{G}_{b}$ & $0$ \\
        $\left[ \texttt{P}_{a},\texttt{P}_{b} \right]$ & $-\epsilon_{ab}\texttt{S}$ & $0$ &$-\epsilon_{ab}\texttt{S}$ & $0$ \\
     \hline
         \end{tabular}
         \captionsetup{font=footnotesize}
    \caption{Commutation relations of the extended Newton-Hooke, extended Bargmann, extended para-Bargmann and extended static algebra.}
    \label{Table7}
\end{table}

\begin{center}
 \begin{figure}[h!]
  \begin{center}
        \includegraphics[width=8.6cm, height=8.1cm]{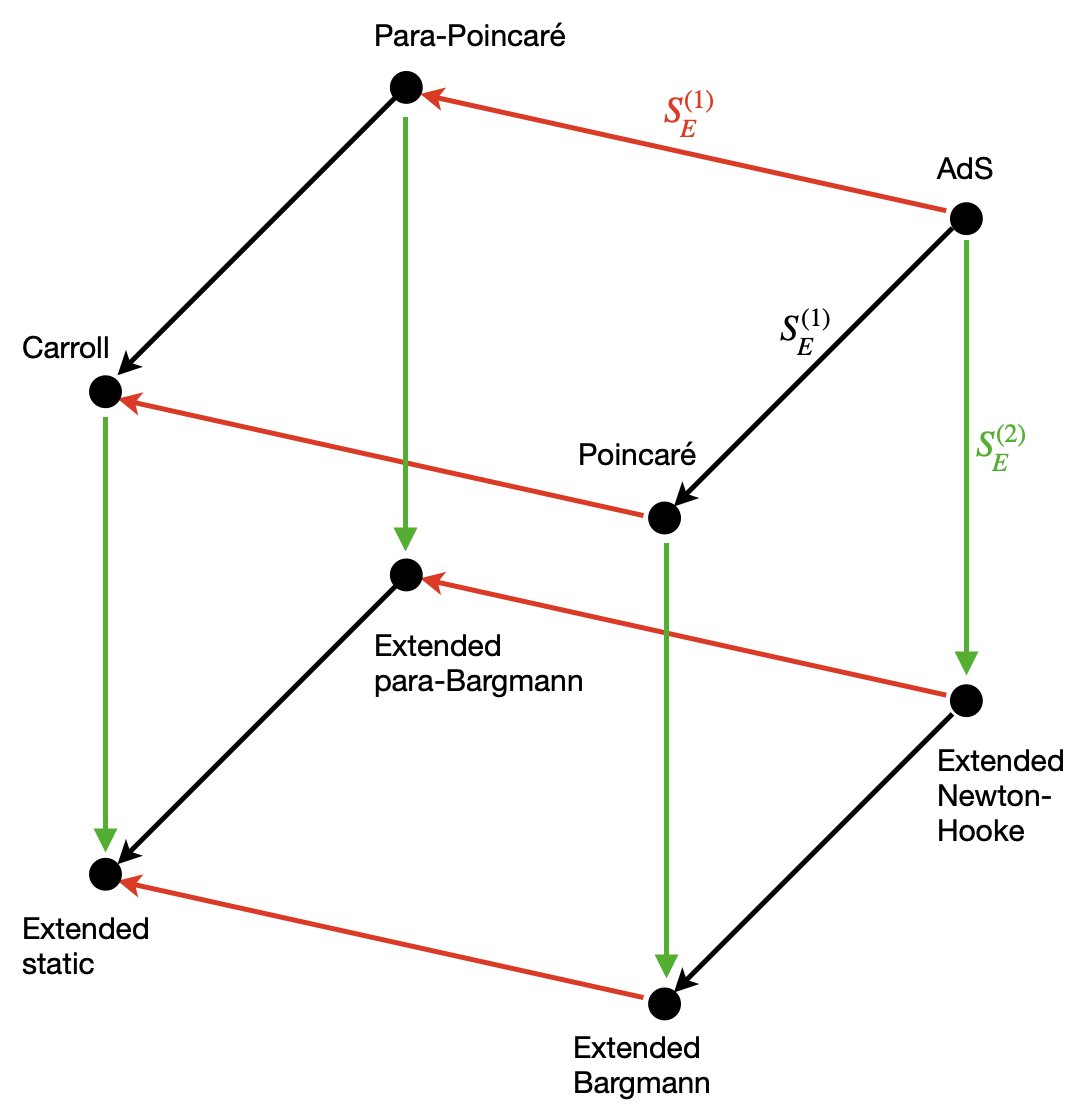}
        \captionsetup{font=footnotesize}
        \caption{This cube summarizes the different non-degenerate $S$-expansion relations starting from the AdS Lie algebra.  There is a space-time resonant $S_{E}^{\left(1\right)}$-expansion (black) and a speed-time resonant $S_{E}^{\left(1\right)}$-expansion (red) which differ in the subspaces decomposition of  $\mathfrak{so}\left(2,2\right)$ (see Table \ref{Table5}). The speed-space resonant $S_{E}^{\left(2\right)}$-expansions (green) reproduce the extended kinematical algebras.}
        \label{fig2}
         \end{center}
        \end{figure}
    \end{center}

The previous results can be visualized in the cube of Figure \ref{fig2}, which summarizes all the sequential expansions starting from the AdS Lie algebra, leading to algebras that admit an invariant metric. Then, the obtained cube extends the original cube of Bacry-Lévy-Leblond \cite{Bacry:1968zf} (see Figure \ref{fig1}) to a new one without degeneracy, for which well-defined CS actions can be constructed. 

Let us note that the cube of Figure \ref{fig2} can be generalized to higher-dimensional spacetime. Nevertheless, it is well known that the Poincaré algebra admits a non-degenerate invariant bilinear trace only in three spacetime dimensions. Remarkably, the resonant $S_{E}^{\left(2\right)}$-expansion of the AdS algebra can also be applied with a Space-time decomposition (see Table \ref{Table5}). In such case, the expanded algebra is characterized by two extra generators $Z_{ab}$ and $Z_{a}$ being related to the $J_{ab}$ and $G_{a}$ AdS generators, respectively \footnote{In tree spacetime dimensions, $Z_{ab}\propto \epsilon_{ab}Z$, $J_{ab}\propto \epsilon_{ab}J$.}, and corresponds to the so-called Maxwell algebra \cite{Bacry:1970du,Bacry:1970ye,Schrader:1972zd,Gomis:2017cmt}, which admits an invariant metric in any spacetime dimension \cite{Soroka:2004fj}. Hence, a cube without degeneracy in higher spacetime dimensions would require to consider the resonant $S_{E}^{\left(2\right)}$-expansion not only with the speed-space decomposition but also the space-time and speed-time ones. In this work, we shall not consider the Maxwell case since we are interested in maintaining the vanishing cosmological constant limit corresponding to the space-time contraction or to a resonant $S_{E}^{\left(1\right)}$-expansion. Nonetheless, it would be interesting to extend the non-degenerate cube to the Maxwell symmetries and analyze if the non-Lorentzian versions of the Maxwell CS gravity theory \cite{Aviles:2018jzw,Concha:2019lhn,Concha:2021jnn} are recovered.
    
\subsection{General kinematical Lie algebras}
%%%%%%%%%%%%%%%%%%%%%%%%%%%%%%%%%%%%%%%%%%%%%%%%%%%%%%%%%%%%%%%%%%%%%%%%%%%%%%%%%%%%%%%%%%%%%%%%%%%%%%%%%%%%%%%%%%%%%%%%%%%%%%%%%%%%%%%%%%%
The non-degenerate cube for extended kinematical algebras previously obtained can be generalized to a family of general kinematical algebras in two different ways. On one hand, a $S_{E}^{\left(N\right)}$-expansion with arbitrary $N$ can be applied to obtain larger kinematical algebras. As we shall see, Post-Newtonian algebras and their Carrollian counterparts are recovered for $N\geq 4$. Nonetheless, analogously to the non-Lorentzian spin-3 symmetries \cite{Concha:2022muu}, non-degeneracy is guaranteed only for even value of $N$.  On the other hand, the $S_{E}^{\left(N\right)}$ semigroup can be applied considering also the speed-time decomposition. In this direction, extended and generalized ultra-relativistic symmetries are found. It is important to emphasize that the $S_{E}^{\left(1\right)}$-expansion in the space-time direction is not modified since we desire to maintain the original relativistic structure given by the AdS and Poincaré algebra. In particular, the space-time $S_{E}^{\left(1\right)}$-expansion can be seen as a vanishing cosmological constant limit between diverse general kinematical algebras.

\subsubsection*{General kinematical algebra and the $S_{E}^{\left(2\right)}$ semigroup}
\begin{center}
 \begin{figure}[h!]
  \begin{center}
        \includegraphics[width=8.9cm, height=9cm]{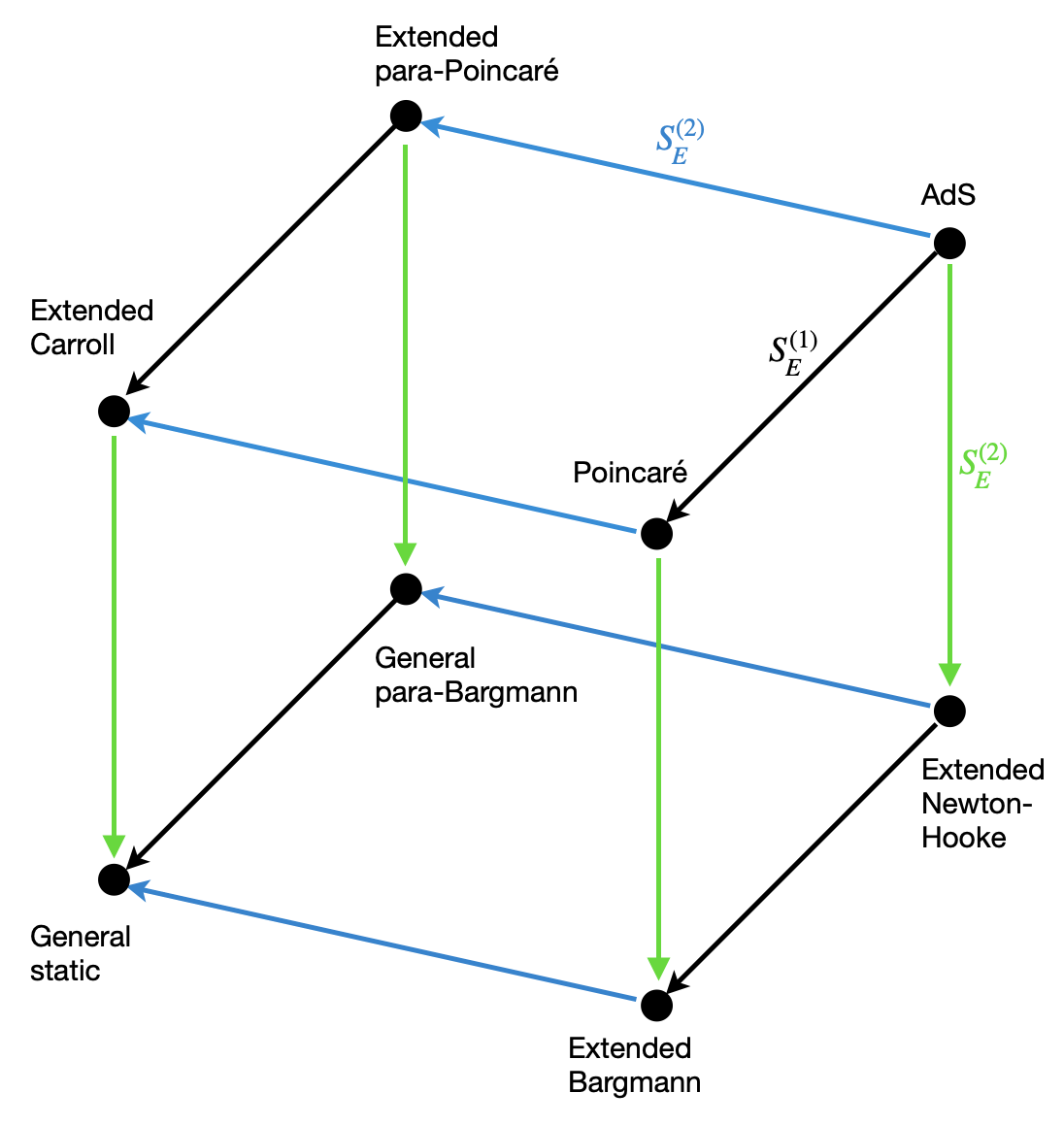}
        \captionsetup{font=footnotesize}
        \caption{This cube summarizes the different $S$-expansion relations starting from the AdS Lie algebra to obtain extended and general kinematical algebras. There is a speed-space resonant $S_{E}^{\left(2\right)}$-expansion (green) and a speed-time resonant $S_{E}^{\left(2\right)}$-expansion (blue). Both speed-space and speed-time expansions are based on the same semigroup $S_{E}^{\left(2\right)}$ but differ in the subspaces decomposition of the original algebra (see Tables \ref{Table8} and \ref{Table10}). The space-time resonant $S_{E}^{\left(1\right)}$-expansions (black) reproduces the usual space-time limit corresponding to the vanishing cosmological constant limit.}
        \label{fig3}
         \end{center}
        \end{figure}
    \end{center}

The simplest generalization of the kinematical Lie algebras can be obtained considering the $S_{E}^{\left(2\right)}$-expansion along the speed-space and speed-time directions as in Figure \ref{fig3}. 
    
To obtain general kinematical algebras considering $S_{E}^{\left(2\right)}$ as the relevant semigroup requires two steps. First, let us consider the speed-space and speed-time decomposition of the AdS (or Poincaré) algebra as in Table \ref{Table4}. Let $S_{E}^{\left(2\right)}$ be the relevant semigroup whose elements satisfy \eqref{mlSE2}. Then, let us consider $S_{E}^{\left(2\right)}=S_0\cup S_1$ as a resonant decomposition given by \eqref{sd2}. Hence, two inequivalent non-Lorentzian algebras appear after applying the resonant $S_{E}^{\left(2\right)}$-expansion to the AdS algebra and considering its $0_S$-reduction. As we have seen in the previous section, the extended Newton-Hooke  and extended Bargmann algebra (see Table \ref{Table7}) are obtained using the speed-space decomposition.
\renewcommand{\arraystretch}{1.1}
\begin{table}[h!]
\centering
    \begin{tabular}{l||C{1,6cm}|C{1,6cm}||C{1,6cm}|C{1,6cm}|}
%\cline{2-2}\cline{3-3}
& \multicolumn{2}{|c||}{Speed-space}& \multicolumn{2}{|c|}{Speed-time} \\ \hline
$\lambda_3$ & \cellcolor[gray]{0.8} & \cellcolor[gray]{0.8} & \cellcolor[gray]{0.8} & \cellcolor[gray]{0.8} \\ \hline
$\lambda_2$ & $\texttt{S}$,\, $\texttt{M}$ & \cellcolor[gray]{0.8} & $\texttt{C}$,\ $\texttt{T}_{a}$ & \cellcolor[gray]{0.8} \\ \hline
$\lambda_1$ & \cellcolor[gray]{0.8} & $\texttt{G}_a$,\ $\texttt{P}_a$ & \cellcolor[gray]{0.8} & $\texttt{H}$,\ $\texttt{G}_{a}$\ \\ \hline
$\lambda_0$ & $ \texttt{J}$,\, $\texttt{H}$ & \cellcolor[gray]{0.8} & $\texttt{J}$,\ $\texttt{P}_{a}$ & \cellcolor[gray]{0.8} \\ \hline
 & $J$, $H$ & $G_a$, $P_a$ & $J$, $P_{a}$ & $H$, $G_{a}$
\end{tabular}
 \captionsetup{font=footnotesize}
\caption{Non-Lorentzian generators in terms of the AdS (or Poincaré) ones and the semigroup elements.}
\label{Table8}%
\end{table}

\noindent A speed-time decomposition as in Table \ref{Table8} allows us to obtain an extended para-Poincaré algebra whose commutation relations appear in Table \ref{Table11}. One can notice that the extended para-Poincaré algebra is isomorphic to the Maxwell Lie algebra \cite{Bacry:1970du,Bacry:1970ye,Schrader:1972zd}, but  physically different\footnote{The Maxwell structure appears by considering the set $\{\texttt{C},\texttt{T}_{a}\}$ as the Maxwellian generators $\{Z,Z_{a}\}$ and by interchanging the role of the generators $\texttt{P}_{a}$ and $\texttt{G}_{a}$.}. On the other hand, the extended Carroll algebra (see Table \ref{Table11}) is obtained starting from the Poincaré algebra instead of the AdS one considering the speed-time decomposition and the resonant $S_{E}^{\left(2\right)}$-expansion.
    
A second step is required to obtain the remaining general kinematical algebras. To this end, there are two inequivalent ways to derive them. Indeed, as one can see from Figure \ref{fig3}, a general para-Bargmann algebra can be obtained by applying the resonant $S_{E}^{\left(2\right)}$-expansion together with its $0_S$-reduction from the extended Para-Poincaré or from the extended Newton-Hooke algebra. Nevertheless, the choice of the original Lie algebra is conditioned to the subspace decomposition. For instance, starting from the extended para-Poincaré algebra requires to consider the speed-space subspaces decomposition of Table \ref{Table9}, while a speed-time subspace decomposition is required starting from the non-relativistic symmetry.
\renewcommand{\arraystretch}{1.1}
\begin{table}[h]
    \centering
    \begin{tabular}{|c||c|c|}
    \hline
       \rowcolor[gray]{0.9} Subspaces  & Speed-space  & Speed-time  \\ 
      \rowcolor[gray]{0.9} & \small{(extended para-Poincaré)} &  \small{(extended Newton-Hooke)} \\ \hline
         $V_0$ & $\texttt{J},\texttt{H},\texttt{C}$& $\texttt{J},\texttt{S},\texttt{P}_{a}$\\
        \rowcolor[gray]{0.9}$V_1$ & $\texttt{G}_{a},\texttt{P}_a,\texttt{T}_{a}$& $\texttt{H},\texttt{M},\texttt{G}_{a}$\\
     \hline
         \end{tabular}
         \captionsetup{font=footnotesize}
    \caption{Subspaces decomposition of extended non-Lorentzian algebras.}
    \label{Table9}
\end{table}

The general para-Bargmann algebra (see Table \ref{Table11}) is characterized by three central charges $\texttt{S},\texttt{M}$ and $\texttt{B}$ which are related to the extended para-Poincaré and extended Newton-Hooke generators through the semigroup elements as in Table \ref{Table10}. The obtained expanded algebra results to be isomorphic to the so-called Maxwellian extended Bargmann (MEB) algebra \cite{Aviles:2018jzw,Concha:2019mxx} but has different physical implications due to the interpretation of the generators $\texttt{P}_a$ and $\texttt{G}_{a}$, which are interchanged with respect to the MEB symmetry. Indeed, the corresponding three-dimensional CS MEB theory is characterized by a vanishing cosmological constant and by the vanishing of the curvatures associated with the spin-connection, $R\left(\omega^{a}\right)=R\left(\omega\right)=0$. On the other hand, as we shall see, the general para-Bargmann gravity theory contains a cosmological constant which acts as a source for the spin-connection curvature $R\left(\omega^{a}\right)=-\epsilon^{ab}{\tau}e_{b}$. Such behavior is due to the presence of the commutator $\left[\texttt{H},\texttt{P}_{a}\right]=\epsilon_{ab}\texttt{G}_{b}$ in the general para-Bargmann symmetry.
\renewcommand{\arraystretch}{1.1}
\begin{table}[h!]
\centering
    \begin{tabular}{l||C{2,0cm}|C{2,0cm}||C{2,0cm}|C{2,0cm}|}
%\cline{2-2}\cline{3-3}
& \multicolumn{2}{|c||}{Speed-space}& \multicolumn{2}{|c|}{Speed-time} \\ \hline
$\lambda_3$ & \cellcolor[gray]{0.8} & \cellcolor[gray]{0.8} & \cellcolor[gray]{0.8} & \cellcolor[gray]{0.8} \\ \hline
$\lambda_2$ & $\texttt{S}$,\quad $\texttt{M}$,\quad $\texttt{B}$ & \cellcolor[gray]{0.8} & $\texttt{C}$,\quad $\texttt{B}$,\quad $\texttt{T}_{a}$ & \cellcolor[gray]{0.8} \\ \hline
$\lambda_1$ & \cellcolor[gray]{0.8} & $\texttt{G}_a$,\ \ $\texttt{P}_a$,\ \ $\texttt{T}_{a}$ & \cellcolor[gray]{0.8} & $\texttt{H}$,\quad $\texttt{M}$,\quad $\texttt{G}_{a}$ \\ \hline
$\lambda_0$ & $ \texttt{J}$,\quad $\texttt{H}$,\quad $\texttt{C}$ & \cellcolor[gray]{0.8} & $\texttt{J}$,\quad $\texttt{S}$,\quad $\texttt{P}_{a}$ & \cellcolor[gray]{0.8} \\ \hline
 & $\texttt{J}$,\quad $\texttt{H}$,\quad $\texttt{C}$ & $\texttt{G}_a$,\ \ $\texttt{P}_a$,\ \ $\texttt{T}_{a}$ & $\texttt{J}$,\quad $\texttt{S}$,\quad $\texttt{P}_{a}$ & $\texttt{H}$,\quad $\texttt{M}$,\quad $\texttt{G}_{a}$
\end{tabular}
\captionsetup{font=footnotesize}
\caption{General kinematical generators in terms of the non-Lorentzian ones and the semigroup elements. The speed-space column contains expanded kinematical generators in terms of the extended para-Poincaré (or extended Carroll) ones. On the other hand, the expanded generators are obtained from the extended Newton-Hooke (or extended Bargmann) ones in the speed-time column.}
\label{Table10}%
\end{table}

A general static algebra (see Table \ref{Table11}) can also be obtained if we consider the resonant $S_{E}^{\left(2\right)}$-expansion and its $0_S$-reduction of the extended Carroll or the extended Bargmann algebra (see Figure \ref{fig3}). Analogously to the general para-Bargmann,  the choice of the starting algebra is conditioned to its subspace decomposition. Indeed, starting from the extended Carroll algebra requires to consider the speed-space decomposition, while a speed-time decomposition is needed if we consider the expansion of the extended Bargmann algebra (see Table \ref{Table10}).
\begin{table}[h]
    \centering
    \begin{tabular}{|l||r|r|r|r|}
    \hline
      \rowcolor[gray]{0.9} Commutators  & Extended  & Extended  & General  & General \\ 
      \rowcolor[gray]{0.9} & para-Poincaré & Carroll & para-Bargmann & static \\ \hline
        $\left[ \texttt{J},\texttt{G}_{a} \right]$ & $\epsilon_{ab}\texttt{G}_{b}$ & $\epsilon_{ab}\texttt{G}_{b}$& $\epsilon_{ab}\texttt{G}_{b}$& $\epsilon_{ab}\texttt{G}_{b}$\\
       \rowcolor[gray]{0.9} $\left[ \texttt{J},\texttt{P}_{a} \right]$ & $\epsilon_{ab}\texttt{P}_{b}$& $\epsilon_{ab}\texttt{P}_{b}$ & $\epsilon_{ab}\texttt{P}_{b}$ & $\epsilon_{ab}\texttt{P}_{b}$ \\
        $\left[ \texttt{G}_{a},\texttt{G}_{b} \right]$ & $-\epsilon_{ab}\texttt{C}$ & $-\epsilon_{ab}\texttt{C}$ &$-\epsilon_{ab}\texttt{B}$ & $-\epsilon_{ab}\texttt{B}$ \\
       \rowcolor[gray]{0.9} $\left[ \texttt{H},\texttt{G}_{a} \right]$ & $\epsilon_{ab}\texttt{T}_{b}$ & $\epsilon_{ab}\texttt{T}_{b}$ &$\epsilon_{ab}\texttt{T}_{b}$ & $\epsilon_{ab}\texttt{T}_{b}$ \\
       $\left[ \texttt{G}_{a},\texttt{P}_{b} \right]$ & $-\epsilon_{ab}\texttt{H}$ & $-\epsilon_{ab}\texttt{H}$ &$-\epsilon_{ab}\texttt{M}$ & $-\epsilon_{ab}\texttt{M}$ \\
       \rowcolor[gray]{0.9} $\left[ \texttt{H},\texttt{P}_{a} \right]$ & $\epsilon_{ab}\texttt{G}_{b}$ & $0$ &$\epsilon_{ab}\texttt{G}_{b}$ & $0$ \\
        $\left[ \texttt{P}_{a},\texttt{P}_{b} \right]$ & $-\epsilon_{ab}\texttt{J}$ & $0$ &$-\epsilon_{ab}\texttt{S}$ & $0$ \\
        \rowcolor[gray]{0.9} $\left[ \texttt{J},\texttt{T}_{a} \right]$ & $\epsilon_{ab}\texttt{T}_{b}$ & $\epsilon_{ab}\texttt{T}_{b}$& $\epsilon_{ab}\texttt{T}_{b}$& $\epsilon_{ab}\texttt{T}_{b}$\\
        $\left[ \texttt{P}_{a},\texttt{T}_{b} \right]$ & $-\epsilon_{ab}\texttt{C}$ & $0$ &$-\epsilon_{ab}\texttt{B}$ & $0$ \\
         \rowcolor[gray]{0.9} $\left[ \texttt{C},\texttt{P}_{a} \right]$ & $\epsilon_{ab}\texttt{T}_{b}$ & $\epsilon_{ab}\texttt{T}_{b}$ &$\epsilon_{ab}\texttt{T}_{b}$ & $\epsilon_{ab}\texttt{T}_{b}$ \\
        
     \hline
         \end{tabular}
         \captionsetup{font=footnotesize}
    \caption{Commutation relations of the extended para-Poincaré, extended Carroll, general para-Bargmann and general static algebra.}
    \label{Table11}
\end{table}

Let us notice that the successive applications of the $S_{E}^{\left(2\right)}$ semigroup generate larger kinematical algebras with additional central charges as it is shown in Table \ref{Table12}. Although some general kinematical algebras coincide in the amount of generators, they are different and cannot be related through a redefinition or change of basis\footnote{The only exception is given by the extended Bargmann and the extended para-Bargmann algebras which turn out to be isomorphic but physically different. Both algebras are related by interchanging the generators $\texttt{P}_{a}$ and $\texttt{G}_{a}$.}.

The cube of Bacry and Lévy-Leblond \cite{Bacry:1968zf} is now generalized through the application of the semigroup $S_{E}^{\left(2\right)}$ along the speed-time direction as in Figure \ref{fig3}. Such generalization might seem unnecessary since the non-degeneracy was already guaranteed for the Para-Poincaré, Carroll, extended para-Bargmann and extended static algebra of the cube in Figure \ref{fig2}. However, the presence of an invariant metric in four or higher spacetime dimensions requires a generalization. For instance, the Para-Poincaré algebra, which is isomorphic to the Poincaré one, does not admit a non-degenerate invariant trace in four spacetime dimensions. The non-degeneracy is restored considering the Maxwell symmetry \cite{Matulich:2019cdo}, which results to be isomorphic to the extended para-Poincaré algebra obtained here as a resonant $S_{E}^{\left(2\right)}$-expansion of the AdS algebra.

\begingroup
\renewcommand{\arraystretch}{1.2}
\begin{table}[h!]
    \centering
    \begin{tabular}{|l||c|c|c|c|c|}
    \hline
      \rowcolor[gray]{0.9}   &  & Extended & Extended  & Extended & General \\ 
      \rowcolor[gray]{0.9} & AdS & Newton-& para- & para- & para-\\ 
      \rowcolor[gray]{0.9} &  & Hooke& Poincaré & Bargmann & Bargmann\\ \hline
        Time generators & $J$, $H$ & $\texttt{J}$, $\texttt{H}$& $\texttt{J}$, $\texttt{H}$, $\texttt{C}$ & $\texttt{J}$, $\texttt{H}$ & $\texttt{J}$, $\texttt{H}$, $\texttt{C}$\\
       \rowcolor[gray]{0.9} Spatial generators & $G_{a}$, $P_{a}$& $\texttt{G}_{a}$, $\texttt{P}_{a}$ & $\texttt{G}_{a}$, $\texttt{P}_{a}$, $\texttt{T}_{a}$ & $\texttt{G}_{a}$, $\texttt{P}_{a}$ & $\texttt{G}_{a}$, $\texttt{P}_{a}$, $\texttt{T}_{a}$\\
       Central charges &  & $\texttt{S}$, $\texttt{M}$ & & $\texttt{S}$, $\texttt{M}$ & $\texttt{S}$, $\texttt{M}$, $\texttt{B}$\\
       \rowcolor[gray]{0.9} Amount of generators & $6$ & $8$ &$9$ & $8$ & $12$\\ \hline \hline
       \rowcolor[gray]{0.9}   & & Extended & Extended  & Extended & General \\ 
      \rowcolor[gray]{0.9} & Poincaré & Bargmann& Carroll & static & static\\ \hline
       Time generators & $J$, $H$ & $\texttt{J}$, $\texttt{H}$ &$\texttt{J}$, $\texttt{H}$, $\texttt{C}$ & $\texttt{J}$, $\texttt{H}$& $\texttt{J}$, $\texttt{H}$, $\texttt{C}$ \\
       \rowcolor[gray]{0.9} Spatial generators & $G_{a}$, $P_{a}$& $\texttt{G}_{a}$, $\texttt{P}_{a}$ & $\texttt{G}_{a}$, $\texttt{P}_{a}$, $\texttt{T}_{a}$ & $\texttt{G}_{a}$, $\texttt{P}_{a}$ & $\texttt{G}_{a}$, $\texttt{P}_{a}$, $\texttt{T}_{a}$\\
        Central charges &  & $\texttt{S}$, $\texttt{M}$ & & $\texttt{S}$, $\texttt{M}$ & $\texttt{S}$, $\texttt{M}$, $\texttt{B}$ \\
        \rowcolor[gray]{0.9} Amount of generators & $6$ & $8$ &$9$ & $8$ & $12$\\
     \hline
         \end{tabular}
         \captionsetup{font=footnotesize}
    \caption{Generators content of the extended and general kinematical algebras.}
    \label{Table12}
\end{table}
\endgroup
\subsubsection*{General kinematical algebra and the $S_{E}^{\left(N\right)}$ semigroup}
The cubes in Figures \ref{fig2} and \ref{fig3} obtained previously can be generalized for a family of non-Lorentzian algebras considering an arbitrary $S_{E}^{\left(N\right)}$ semigroup. Our purpose is to present the most general scheme involving non-Lorentzian symmetries. As we shall see, known Newtonian, post-Newtonian algebras and their Carrollian counterparts can be obtained for particular values of $N$.

Let us consider first $S_{E}^{\left(N\right)}=\{\lambda_0,\lambda_1,\lambda_2,\cdots,\lambda_{N+1}\}$ as the relevant semigroup whose elements satisfy
\begin{equation}
\lambda _{\alpha }\lambda _{\beta }=\left\{ 
\begin{array}{lcl}
\lambda _{\alpha +\beta }\,\,\,\, & \mathrm{if}\,\,\,\,\alpha +\beta \leq
N+1\,, &  \\ 
\lambda _{N+1}\,\,\, & \mathrm{if}\,\,\,\,\alpha +\beta >N+1\,, & 
\end{array}%
\right.   \label{mlSEN}
\end{equation}%
with $\lambda_{N+1}=0_S$ being the zero element of the semigroup. Let $S_{E}^{\left(N\right)}=S_{0}\cup S_{1}$ be a subset decomposition of the semigroup with
\begin{eqnarray}
S_{0} &=&\left\{ \lambda _{2m},\ \text{with }m=0,\ldots ,\left[ \frac{N}{2}%
\right] \right\} \cup \{\lambda _{N+1}\}\,, \notag \\
S_{1} &=&\left\{ \lambda _{2m+1},\ \text{with }m=0,\ldots ,\left[ \frac{N+1}{%
2}\right] \right\} \cup \{\lambda _{N+1}\}\,, \label{sdN}
\end{eqnarray}%
where $[\ldots ]$ denotes the integer part. The subset decomposition \eqref{sdN} is said to be resonant since it satisfies the same algebraic structure than the subspace decomposition \eqref{sd} of the original $\mathfrak{so}\left(2,2\right)$ algebra (or Poincaré). Then, generalized kinematical Lie algebras are obtained after considering a resonant $S_{E}^{\left(N\right)}$-expansion of the AdS algebra (or Poincaré) and applying a $0_S$-reduction. In particular, the expanded generators are expressed in terms of the original ones as in Table \ref{Table13}. 
\begingroup
\renewcommand{\arraystretch}{1.25}
\begin{table}[h!]
    \centering
    \begin{tabular}{|l||c|c||l|c|c|c|}
    \hline
      \rowcolor[gray]{0.9}  Expanded  & $\mathfrak{nh}^{\left(N\right)}$  & $\mathfrak{pp}^{\left(N\right)}$ & Expanded& $\mathfrak{pg}^{\left(N\right)}$ & $\mathfrak{pg}^{\left(N\right)}$ \\ 
      \rowcolor[gray]{0.9} generators & from AdS & from AdS & generators & from $\mathfrak{nh}^{\left(N\right)}$& from $\mathfrak{pp}^{\left(N\right)}$ \\ 
        $\texttt{J}^{\left(m\right)}$ & $\lambda_{2m} J$ & $\lambda_{2m} J$ & $\texttt{J}^{\left(n,m\right)}$&$\lambda_{2m} \texttt{J}^{\left(n\right)}$ & $\lambda_{2m} \texttt{J}^{\left(n\right)}$ \\
       \rowcolor[gray]{0.9} $\texttt{H}^{\left(m\right)}$ & $\lambda_{2m} H$& $\lambda_{2m+1} H$ & $\texttt{H}^{\left(n,m\right)}$ & $\lambda_{2m+1} \texttt{H}^{\left(n\right)}$ & $\lambda_{2m} \texttt{H}^{\left(n\right)}$ \\
       $\texttt{G}_{a}^{\left(m\right)}$ & $\lambda_{2m+1}G_{a}$ & $\lambda_{2m+1}G_{a}$ & $\texttt{G}_{a}^{\left(n,m\right)}$ & $\lambda_{2m+1}\texttt{G}_{a}^{\left(n\right)}$& $\lambda_{2m+1}\texttt{G}_{a}^{\left(n\right)}$ \\
       \rowcolor[gray]{0.9} $\texttt{P}_{a}^{\left(m\right)}$ & $\lambda_{2m+1}P_{a}$ & $\lambda_{2m}P_{a}$ & $\texttt{P}_{a}^{\left(n,m\right)}$ &$\lambda_{2m}\texttt{P}_{a}^{\left(n\right)}$ & $\lambda_{2m+1}\texttt{P}_{a}^{\left(n\right)}$\\ \hline 
         \end{tabular}
         \captionsetup{font=footnotesize}
    \caption{Expanded generators in terms of the original ones and the semigroup elements.}
    \label{Table13}
\end{table}
\endgroup
\begin{center}
 \begin{figure}[h!]
  \begin{center}
        \includegraphics[width=8.2cm, height=8.2cm]{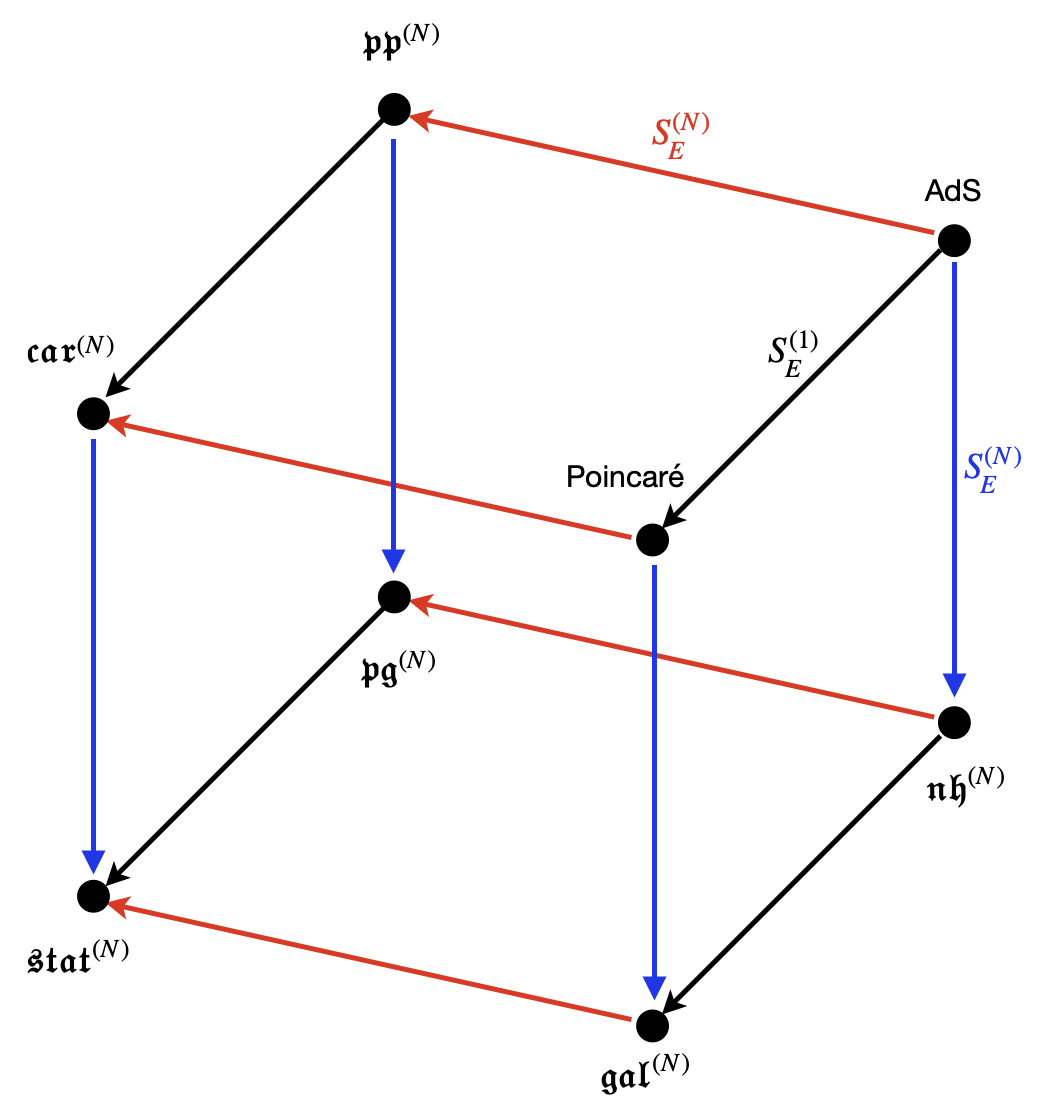}
        \captionsetup{font=footnotesize}
        \caption{This cube summarizes the different $S_{E}^{\left(N\right)}$-expansion relations starting from the AdS Lie algebra to obtain general kinematical algebras.}
        \label{fig4}
         \end{center}
        \end{figure}
    \end{center}
    
In the ultra-relativistic regime we obtain generalizations of the para-Poincaré and Carroll algebras (see Table \ref{Table14}) which we have denoted as $\mathfrak{pp}^{\left(N\right)}$ and $\mathfrak{car}^{\left(N\right)}$, respectively. In the non-relativistic counterpart,  we obtain the generalized Newton-Hooke and Galilean algebras \cite{Gomis:2019nih} which have been denoted as $\mathfrak{nh}^{\left(N\right)}$ and $\mathfrak{gal}^{\left(N\right)}$, respectively.

Two other generalized kinematical algebras, denoted as $\mathfrak{stat}^{\left(N\right)}$ and $\mathfrak{pg}^{\left(N\right)}$ (see Table \ref{Table15}), appear by applying the resonant $S_{E}^{\left(N\right)}$-expansion to the generalized non-Lorentzian symmetries together with the $0_S$-reduction condition (see Figure \ref{fig4}). The relations between the expanded generators and the original ones depend on the nature of the starting algebra. Indeed, the expanded generators defined in terms of the ultra-relativistic kinematical generators are defined quite differently from those derived from the non-relativistic ones (see Table \ref{Table13}). For instance, the $\mathfrak{pg}^{\left(2\right)}$ algebra, corresponding to the general para-Bargmann, contains $S$ as a central charge which corresponds to $\texttt{J}^{\left(1,0\right)}$ if it is obtained from the $\mathfrak{nh}^{\left(2\right)}$ algebra. On the other hand, the central charge $S$ coincides with $\texttt{J}^{\left(0,1\right)}$ if it is derived by expanding the $\mathfrak{pp}^{\left(2\right)}$ algebra. The upper index $\left(n,m\right)$ in the expanded generators of Table \ref{Table13} reflects the application of two successive expansions: the first applied to the relativistic AdS (or Poincaré) algebra and the second to its non-Lorentzian counterpart.

\begin{table}[h!]
\renewcommand{\arraystretch}{1.85}
    \centering
    \begin{tabular}{|l||r|r|r|r|}
    \hline
      \rowcolor[gray]{0.9} Commutators  & $\mathfrak{nh}^{\left(N\right)}$ & $\mathfrak{gal}^{\left(N\right)}$ & $\mathfrak{pp}^{\left(N\right)}$  & $\mathfrak{car}^{\left(N\right)}$   
      \\ \hline
        $\left[ \texttt{J}^{(n)},\texttt{G}_{a}^{(m)} \right]$ & $\epsilon_{ab}\texttt{G}_{b}^{(n+m)}$ & $\epsilon_{ab}\texttt{G}_{b}^{(n+m)}$& $\epsilon_{ab}\texttt{G}_{b}^{(n+m)}$& $\epsilon_{ab}\texttt{G}_{b}^{(n+m)}$  \\
       \rowcolor[gray]{0.9} $\left[ \texttt{J}^{(n)},\texttt{P}_{a}^{(m)} \right]$ & $\epsilon_{ab}\texttt{P}_{b}^{(n+m)}$& $\epsilon_{ab}\texttt{P}_{b}^{(n+m)}$ & $\epsilon_{ab}\texttt{P}_{b}^{(n+m)}$ & $\epsilon_{ab}\texttt{P}_{b}^{(n+m)}$  \\
        $\left[ \texttt{G}_{a}^{(n)},\texttt{G}_{b}^{(m)} \right]$ & $-\epsilon_{ab}\texttt{J}^{(n+m+1)}$ & $-\epsilon_{ab}\texttt{J}^{(n+m+1)}$ &$-\epsilon_{ab}\texttt{J}^{(n+m+1)}$ & $-\epsilon_{ab}\texttt{J}^{(n+m+1)}$ \\
       \rowcolor[gray]{0.9} $\left[ \texttt{H}^{(n)},\texttt{G}_{a}^{(m)} \right]$ & $\epsilon_{ab}\texttt{P}_{b}^{(n+m)}$ & $\epsilon_{ab}\texttt{P}_{b}^{(n+m)}$ &$\epsilon_{ab}\texttt{P}_{b}^{(n+m+1)}$ & $\epsilon_{ab}\texttt{P}_{b}^{(n+m+1)}$ \\
       $\left[ \texttt{G}_{a}^{(n)},\texttt{P}_{b}^{(m)} \right]$ & $-\epsilon_{ab}\texttt{H}^{(n+m+1)}$ & $-\epsilon_{ab}\texttt{H}^{(n+m+1)}$ & $-\epsilon_{ab}\texttt{H}^{(n+m)}$ & $-\epsilon_{ab}\texttt{H}^{(n+m)}$ \\
       \rowcolor[gray]{0.9} $\left[ \texttt{H}^{(n)},\texttt{P}_{a}^{(m)} \right]$ & $\epsilon_{ab}\texttt{G}_{b}^{(n+m)}$ & $0$ &$\epsilon_{ab}\texttt{G}_{b}^{(n+m)}$ & $0$ \\
        $\left[ \texttt{P}_{a}^{(n)},\texttt{P}_{b}^{(m)} \right]$ & $-\epsilon_{ab}\texttt{J}^{(n+m+1)}$ & $0$ &$-\epsilon_{ab}\texttt{J}^{(n+m)}$ & $0$ \\
     \hline
         \end{tabular}
         \captionsetup{font=footnotesize}
    \caption{Commutation relations of the non-Lorentzian kinematical Lie algebras.}
    \label{Table14}
\end{table}

Let us note that the original cube of Bacry and Lévy-Leblond \cite{Bacry:1968zf} is recovered for $N=1$ in which the resonant $S_{E}^{\left(1\right)}$-expansion reproduces the different limits starting from the $\mathfrak{so}\left(2,2\right)$ Lie algebra (see Figure \ref{fig1}). For $N=2$, we recover the generalized cube presented in Figure \ref{fig3} which is characterized by a non-degenerate invariant trace in the non-relativistic regime. Interestingly, for $N=3$ the so-called Newtonian algebra is obtained as a resonant $S_{E}^{\left(3\right)}$-expansion of the Poincaré algebra. The Newtonian algebra appears as the underlying symmetry of an action principle for Newtonian gravity \cite{Hansen:2019pkl}. Such symmetry can be seen as an extension of the extended Bargmann algebra and is characterized by the presence of additional generators $\{\texttt{B}_{a},\texttt{T}_{a}\}$ which coincides with our $\texttt{G}_{a}^{\left(3\right)}$ and $\texttt{P}_{a}^{\left(3\right)}$ generators of Table \ref{Table13}. A Newton-Hooke version of the Newtonian algebra \cite{Concha:2022jdc} is obtained if we apply the $0_S$-reduced resonant $S_{E}^{\left(3\right)}$-expansion to the AdS algebra, meanwhile Carrollian versions of the Newtonian symmetries appear in the ultra-relativistic regime. The inconvenient with $N=3$, analogously to the $N=1$ case, is the degeneracy of the invariant tensor for the kinematical algebras obtained in the non-relativistic direction. Such degeneracy can be avoided considering $N=4$ which reproduces the extended Newtonian algebra  \cite{Ozdemir:2019orp} and its Newton-Hooke version \cite{Concha:2019dqs,Concha:2021jos}. For $N>4$, the expanded kinematical algebras reproduce post-Newtonian algebras together with their Carrollian versions.
\begin{table}[h]
\renewcommand{\arraystretch}{1.85}
    \centering
    \begin{tabular}{|l||r|r|}
    \hline
      \rowcolor[gray]{0.9} Commutators  & $\mathfrak{pg}^{\left(N\right)}$ & $\mathfrak{stat}^{\left(N\right)}$ 
      \\ \hline
        $\left[ \texttt{J}^{(n,p)},\texttt{G}_{a}^{(m,q)} \right]$ & $\epsilon_{ab}\texttt{G}_{b}^{(n+m,p+q)}$ & $\epsilon_{ab}\texttt{G}_{b}^{(n+m,p+q)}$  \\
       \rowcolor[gray]{0.9} $\left[ \texttt{J}^{(n,p)},\texttt{P}_{a}^{(m,q)} \right]$ & $\epsilon_{ab}\texttt{P}_{b}^{(n+m,p+q)}$& $\epsilon_{ab}\texttt{P}_{b}^{(n+m,p+q)}$  \\
        $\left[ \texttt{G}_{a}^{(n,p)},\texttt{G}_{b}^{(m,q)} \right]$ & $-\epsilon_{ab}\texttt{J}^{(n+m+1,p+q+1)}$ & $-\epsilon_{ab}\texttt{J}^{(n+m+1,p+q+1)}$  \\
       \rowcolor[gray]{0.9} $\left[ \texttt{H}^{(n,p)},\texttt{G}_{a}^{(m,q)} \right]$ & $\epsilon_{ab}\texttt{P}_{b}^{(n+m,p+q+1)}$ & $\epsilon_{ab}\texttt{P}_{b}^{(n+m,p+q+1)}$  \\
       $\left[ \texttt{G}_{a}^{(n,p)},\texttt{P}_{b}^{(m,q)} \right]$ & $-\epsilon_{ab}\texttt{H}^{(n+m+1,p+q)}$ & $-\epsilon_{ab}\texttt{H}^{(n+m+1,p+q)}$  \\
       \rowcolor[gray]{0.9} $\left[ \texttt{H}^{(n,p)},\texttt{P}_{a}^{(m,q)} \right]$ & $\epsilon_{ab}\texttt{G}_{b}^{(n+m,p+q)}$ & $0$  \\
        $\left[ \texttt{P}_{a}^{(n,p)},\texttt{P}_{b}^{(m,q)} \right]$ & $-\epsilon_{ab}\texttt{J}^{(n+m+1,p+q)}$ & 0  \\
     \hline
         \end{tabular}
         \captionsetup{font=footnotesize}
    \caption{Commutation relations of the generalized kinematical Lie algebras $\mathfrak{pg}^{\left(N\right)}$ and $\mathfrak{stat}^{\left(N\right)}$ by expanding the $\mathfrak{nh}^{\left(N\right)}$ and the $\mathfrak{gal}^{\left(N\right)}$ algebra, respectively. The upper index $(n,p)$ in the generators denotes two sub-sequential $S$-expansions. }
    \label{Table15}
\end{table}
%%%%%%%%%%%%%%%%%%%%%%%%%%%%%%%%%%%%%%%%%%%%%%%%%%%%%%%%%%%%%%%%%%%%%%%
%%%%%%%%%%%%%%%%%%%%%%%%%%%%%%%%%%%%%%%%%%%%%%%%%%%%%%%%%%%%%%%%%%%%%%%%%%%%
\section{Extended kinematical Chern-Simons gravity theories in three spacetime dimensions}\label{sec4}

In this section we apply the $S$-expansion method to derive the diverse CS actions associated with the extended kinematical algebras obtained in the cube of Figure \ref{fig2}. The choice of the cube is not arbitrary but is due to the fact that the extended kinematical algebras, unlike the usual kinematical algebras presented in \cite{Bacry:1968zf}, are characterized by the fact that they admit a non-degenerate invariant bilinear trace ensuring the construction of a well-defined CS action. As we shall see, each kinematical CS action can be obtained as sequential expansions starting from the AdS CS one by expressing the expanded gauge fields in terms of the original ones. In particular, the $\mathfrak{so}\left(2,2\right)$ Lie algebra admits the following non-vanishing components of the invariant tensor:
\begin{align}
    \langle J J \rangle &= -\mu_0 \,, & \langle J  H \rangle &= -\mu_1\,, & \langle H H \rangle&=-\mu_0\,,\notag \\
    \langle G_a G_b \rangle &=\mu_0\delta_{ab}\,, & \langle G_a P_b\rangle &=\mu_1\delta_{ab}\,, &\langle P_a P_b \rangle&=\mu_0\delta_{ab}\,, \label{IT}
\end{align}
where $\mu_0$ is related to an exotic sector of the three-dimensional AdS CS gravity \cite{Witten:1988hc,Zanelli:2005sa}. Let us note that in the Poincaré limit we have $\langle H H \rangle = \langle P_a P_b \rangle =0$. 

According to Theorem VII of \cite{Izaurieta:2006zz}, the $S$-expansion procedure has the advantage of providing us with the invariant tensor of the expanded algebra in terms of the original one. Remarkably, the expanded algebras obtained in the cube of Figure \ref{fig2} admit non-degenerate invariant tensors. In particular, the invariant tensor for each extended kinematical algebra is obtained considering the resonant expansions employed to obtain the different symmetry algebras of the cube in Figure \ref{fig2}. The explicit non-vanishing components of the invariant tensors for each expanded algebra are listed in Table \ref{Table16}, in which the ultra-relativistic generators are related to the relativistic ones through the semigroup elements as in Table \ref{Table5}. On the other hand, the generators involved in the invariant tensors for the extended kinematical algebra are related to the relativistic and ultra-relativistic ones as in Table \ref{Table6}. 
\begingroup
\renewcommand{\arraystretch}{1.13}
\begin{table}[h!]
    \centering
    \begin{tabular}{|l||c|c|c|c|c|c|}
    \hline
      \rowcolor[gray]{0.9}    & Para- &   & Extended & Extended & Extended & Extended \\ 
      \rowcolor[gray]{0.9}  & Poincaré& Carroll & Newton- & Bargmann & para- & static \\ 
      \rowcolor[gray]{0.9}   & &  & Hooke & &Bargmann & \\ \hline
      $\langle \texttt{J} \texttt{J} \rangle$ & $-\beta_0$ & $-\beta_0$ & $0$ & $0$ & $0$ & $0$ \\
       \rowcolor[gray]{0.9} $\langle \texttt{J} \texttt{H} \rangle$ & $-\beta_1$ & $-\beta_1$ & $0$ & $0$ & $0$ & $0$ \\
       $\langle \texttt{H} \texttt{H} \rangle$ & $0$ & $0$ & $0$ & $0$ & $0$& $0$ \\
        \rowcolor[gray]{0.9}$\langle \texttt{G}_{a} \texttt{G}_{b} \rangle$ & $0$& $0$ & $\alpha_0 \delta_{ab}$ & $\alpha_0 \delta_{ab}$ & $0$ & $0$ \\
       $\langle \texttt{G}_{a} \texttt{P}_{b} \rangle$ & $\beta_1 \delta_{ab}$ & $\beta_1 \delta_{ab}$ & $\alpha_1 \delta_{ab}$ & $\alpha_1 \delta_{ab}$ & $\gamma_1 \delta_{ab}$ & $\gamma_1 \delta_{ab}$ \\
         \rowcolor[gray]{0.9}  $\langle \texttt{P}_{a} \texttt{P}_{b} \rangle$ & $\beta_0 \delta_{ab}$ & $0$ & $\alpha_0 \delta_{ab}$ & $0$ & $\gamma_0 \delta_{ab}$ & $0$\\ \hline 
         $\langle \texttt{J} \texttt{S} \rangle$ & - & - & $-\alpha_0$ & $-\alpha_0$ & $-\gamma_0$ & $-\gamma_0$ \\
         \rowcolor[gray]{0.9} $\langle \texttt{J} \texttt{M} \rangle$ & - & - & $-\alpha_1$ & $-\alpha_1$ & $-\gamma_1$ & $-\gamma_1$ \\
         $\langle \texttt{H} \texttt{M} \rangle$ & - & - & $-\alpha_0$ & $0$ & $0$ & $0$ \\ 
         \rowcolor[gray]{0.9} $\langle \texttt{H} \texttt{S} \rangle$ & - & - & $-\alpha_1$ & $-\alpha_1$ & $-\gamma_1$ & $-\gamma_1$ \\ 
         \hline
         \end{tabular}
         \captionsetup{font=footnotesize}
    \caption{Non-vanishing components of the invariant tensor for the expanded kinematical algebras.}
    \label{Table16}
\end{table}
\endgroup

\noindent Let us note that the constants appearing in the expanded invariant tensor are related to the original ones as follows:
\begin{align}
    \beta_0&=\lambda_0 \mu_0\,, &\beta_1&=\lambda_1 \mu_1\,, \notag \\
    \alpha_0&=\lambda_2 \mu_0\,, &\alpha_1&=\lambda_2 \mu_1\,, \notag \\
    \gamma_0&=\lambda_2 \beta_0=\lambda_0\alpha_0\,, &\gamma_1&=\lambda_2 \beta_1=\lambda_1\alpha_1\,, \label{const}
\end{align}
where the $\gamma$'s can be obtained from either the ultra-relativistic constants or the non-relativistic ones. Nevertheless, starting from a non-relativistic algebra requires to consider $S_{E}^{\left(1\right)}$ as the relevant semigroup with $\lambda_2=0_S$ being the zero element (see the cube in Figure \ref{fig2}). 

A CS action invariant under the kinematical Lie algebras of the cube in Figure \ref{fig2} can be constructed  by considering the non-vanishing components of the invariant tensor given by \eqref{IT} and Table \ref{Table16} together with the corresponding gauge connection one-form $A=A^{A}T_{A}$ in the general expression of the CS action,
\begin{eqnarray}
I_{CS}=\frac{k}{4\pi}\int\langle AdA+\frac{2}{3}A^3\rangle\,.\label{CS}
\end{eqnarray}
Here $k$ denotes the CS level of the theory which is related to the gravitational constant $G$ through $k=1/(4G)$. In the relativistic case, the gauge connection one-form $A$ reads
\begin{align}
    A&=W J+V H+W^{a}G_{a}+V^{a}P_{a}\,.\label{relA}
\end{align}
Here $W$ represent the spin-connection for boosts, $W_{a}$ is the spatial spin-connection, $V$ is the time-like vielbein and $V^{a}$ corresponds to the spatial vielbein. The curvature two-form $F$ is given by
\begin{align}
F=&R\left(W\right)J+R\left(V\right)H+R^{a}\left(W^{b}\right)G_{a}+R^{a}\left(V^{b}\right)P_{a}\,, \label{relF}
\end{align}
with
\begin{eqnarray}
R\left(W\right)&=&dW+\frac{1}{2}\epsilon^{ac}W_a W_c+\frac{1}{2\ell^2}\epsilon^{ac}V_{a}V_{c}\,, \notag \\
R\left(V\right)&=&dV + \epsilon^{ac}W_{a}V_{c}\,, \notag \\
R^{a}\left(W^{b}\right)&=&dW^{a}+\epsilon^{ac}W W_c+\frac{1}{\ell^2}\epsilon^{ac}V V_c\,, \notag\\
R^{a}\left(V^{b}\right)&=&dV^{a}+\epsilon^{ac}W V_c+\epsilon^{ac}V W_c\,. \label{relcurv}
\end{eqnarray}
Here, we have considered the space-time rescaling of the generators as in Table \ref{Table1} in order to introduce explicitly the  cosmological constant  through $\Lambda = - \frac{1}{\ell^2}$. Then, in the vanishing cosmological constant limit $\ell\rightarrow\infty$ we recover the Poincaré curvature two-forms. 

The gauge connection one-form for the ultra-relativistic algebras is given by
\begin{align}
A&=\omega\texttt{J}+\tau\texttt{H}+\omega^{a}\texttt{G}_{a}+e^{a}\texttt{P}_{a}\,.\label{ultA}
\end{align}
In particular, the curvature two-form for the para-Poincaré symmetry algebra reads
\begin{align}
F&=R\left(\omega\right)\texttt{J}+R\left(\tau\right)\texttt{H}+R^{a}\left(\omega^{b}\right)\texttt{G}_{a}+R^{a}\left(e^{b}\right)\texttt{P}_{a}\,,\label{ultF}
\end{align}
where the components are defined in Table \ref{Table17}. 
\begin{table}[h!]
\renewcommand{\arraystretch}{1.4}
    \centering
    \begin{tabular}{|l||c|c|c|}
    \hline
      \rowcolor[gray]{0.9}  & Para-Poincaré & Extended Newton-Hooke & Extended para-Bargmann
      \\ \hline
        $R\left(\omega\right)$ & $d\omega+\frac{1}{2\ell^2}\epsilon^{ac}e_{a}e_{c}$ & $d\omega$ & $d\omega$  \\
       \rowcolor[gray]{0.9} $R\left(\tau\right)$ & $d\tau + \epsilon^{ac}\omega_{a}e_{c}$ & $d\tau$ & $d\tau$  \\
        $R^{a}\left(\omega^{b}\right)$ & $d\omega^{a}+\epsilon^{ac}\omega \omega_c+\frac{1}{\ell^2}\epsilon^{ac}\tau e_c$ & $d\omega^{a}+\epsilon^{ac}\omega \omega_c+\frac{1}{\ell^2}\epsilon^{ac}\tau e_c$ & $d\omega^{a}+\epsilon^{ac}\omega \omega_c+\frac{1}{\ell^2}\epsilon^{ac}\tau e_c$   \\
       \rowcolor[gray]{0.9} $R^{a}\left(e^{b}\right)$ & $de^{a}+\epsilon^{ac}\omega e_c$ & $de^{a}+\epsilon^{ac}\omega e_c+\epsilon^{ac}\tau\omega_c$  & $de^{a}+\epsilon^{ac}\omega e_c$  \\
       $R^{a}\left(s\right)$ & - & $ds+\frac{1}{2}\epsilon^{ac}\omega_a\omega_c+\frac{1}{2\ell^2}\epsilon^{ac}e_{a}e_{c}$  & $ds+\frac{1}{2\ell^2}\epsilon^{ac}e_{a}e_{c}$  \\
       \rowcolor[gray]{0.9} $R^{a}\left(m\right)$ & - & $dm+\epsilon^{ac}\omega_{a}e_{c}$ & $dm+\epsilon^{ac}\omega_{a}e_{c}$ \\
     \hline
         \end{tabular}
         \captionsetup{font=footnotesize}
    \caption{Curvature two-forms for the Para-Poincaré, extended Newton-Hooke and extended para-Bargmann algebras. The flat limit $\ell\rightarrow\infty$ reproduces the curvatures for the Carroll, extended Bargmann and extended static algebras.}
    \label{Table17}
\end{table}

As in the relativistic case, we have introduced the $\ell$ parameter through the space-time rescaling considered in Table \ref{Table1}. In the vanishing cosmological constant limit $\ell\rightarrow\infty$ we obtain the Carroll curvature two-forms. On the other hand, the gauge connection one-form taking values in the extended kinematical Lie algebra reads
\begin{align}
A&=\omega\texttt{J}+\tau\texttt{H}+\omega^{a}\texttt{G}_{a}+e^{a}\texttt{P}_{a}+s\texttt{S}+m\texttt{M}\,,\label{extA}
\end{align}
where $s$ and $m$ are the gauge fields associated to the central charges $\texttt{S}$ and $\texttt{M}$, respectively. Although the extended kinematical algebras share the same expression for the curvature two-form $F$, they differ in its components. Indeed, the curvature two-form for the extended kinematical Lie algebras is
\begin{align}
F&=R\left(\omega\right)\texttt{J}+R\left(\tau\right)\texttt{H}+R^{a}\left(\omega^{b}\right)\texttt{G}_{a}+R^{a}\left(e^{b}\right)\texttt{P}_{a}+R^{a}\left(s\right)\texttt{S}+R^{a}\left(m\right)\texttt{M}\,,\label{extF}
\end{align}
where the components are given in Table \ref{Table17}.

The relativistic $\mathfrak{so}\left(2,2\right)$ three-dimensional CS gravity action is obtained using the connection 1-form \eqref{relA} and the non-vanishing components of the invariant
tensor \eqref{IT} in the general expression \eqref{CS},
\begin{eqnarray}
    I_{CS}=\frac{k}{4\pi}\int \biggl\{ &\mu_0& \left[-WdW+\frac{1}{\ell^2}V_{a}R^{a}\left(V^{b}\right)-\frac{1}{\ell^2}VdV+W_{a}R^{a}\left(W^{b}\right) \right] \notag \\
    &+& \mu_1 \left[-2WdV+W_{a}R^{a}\left(V^{b}\right)+V_{a}R^{a}\left(W^{b}\right) \right] \biggr\} \,, \label{Iads}
\end{eqnarray}
where the curvature two-forms are defined in \eqref{relcurv}. Naturally, we recover the $\mathfrak{iso}\left(2,1\right)$ CS action in the flat limit $\ell\rightarrow\infty$. 

The CS action for the ultra-relativistic and the extended kinematical Lie algebras can be constructed  by considering the respective gauge connection one-form \eqref{ultA} and \eqref{extA} together with the corresponding non-vanishing components of the invariant tensor. The explicit terms appearing in the corresponding CS action are listed in Table \ref{Table18}. In particular, each term appears in a particular sector of the action according to Table \ref{Table16} and is expressed in terms of the curvature two-forms defined in Table \ref{Table17}. 
\begin{table}[h!]
\renewcommand{\arraystretch}{1.4}
    \centering
    \begin{tabular}{|l||c|c|c|}
    \hline
      \rowcolor[gray]{0.9} CS terms & Para-Poincaré & Extended  & Extended 
      \\
      \rowcolor[gray]{0.9}  &  & Newton-Hooke  & para-Bargmann
      \\
      \hline
        $-\omega d\omega$ & $\beta_0$ & - & -  \\
       \rowcolor[gray]{0.9} $\frac{1}{\ell^2}e_{a}R^{a}\left(e^{b}\right)$ & $\beta_0$ & $\alpha_0$ & $\gamma_0$  \\
        $\omega_aR^{a}\left(\omega^{b}\right)$ & - & $\alpha_0$ & -   \\
       \rowcolor[gray]{0.9} $-2sR\left(\omega\right)$ & - & $\alpha_0$  & $\gamma_0$  \\
       $-\frac{2}{\ell^2}mR\left(\tau\right)$ & - & $\alpha_0$  & -  \\
       \rowcolor[gray]{0.9} $-2\omega d\tau$ & $\beta_1$ & - & - \\
       $\omega_a R^{a}\left(e^{b}\right)$ & $\beta_1$ & $\alpha_1$ & $\gamma_1$ \\
        \rowcolor[gray]{0.9} $e_{a}R^{a}\left(\omega^{b}\right)$ & $\beta_1$ & $\alpha_1$ & $\gamma_1$ \\
        $-2sR\left(\tau\right)$ & - & $\alpha_1$ & $\gamma_1$ \\
        \rowcolor[gray]{0.9} $-2mR\left(\omega\right)$ & - & $\alpha_1$ & $\gamma_1$ \\
     \hline
         \end{tabular}
         \captionsetup{font=footnotesize}
    \caption{List of the CS terms appearing in the ultra-relativistic and extended kinematical gravity theories. The constants indicate in which sector of the theory the CS terms are present. The explicit expressions for the curvature two-forms for a given kinematical Lie algebra appear in Table \ref{Table17}.}
    \label{Table18}
\end{table}

The term proportional to $\beta_0$ corresponds to an exotic ultra-relativistic action, while $\beta_1$ reproduces the para-Poincaré gravity action. The latter results to be isomorphic to the relativistic Poincaré CS action by interchanging the spatial spin-connection and the spatial vielbein. In the vanishing cosmological constant limit $\ell\rightarrow\infty$, the terms proportional to $\alpha$'s reproduce the most general Carroll gravity action. On the other hand, along $\alpha_0$ and $\alpha_1$ we obtain an extended Newton-Hooke gravity action which is known to describe a non-relativistic model with cosmological constant. The extended Newton-Hooke algebra can alternatively be obtained as a non-relativistic limit of the $\mathfrak{so}\left(2,2\right)\oplus\mathfrak{u}\left(1\right)^{2}$ algebra. In our procedure, no $U\left(1\right)$ enlargement is required to get the desired central extensions. Here, the flat limit $\ell\rightarrow\infty$ reproduces the extended Bargmann CS gravity action \cite{Papageorgiou:2009zc,Bergshoeff:2016lwr}. Finally, the CS gravity action for the extended para-Bargmann algebra and its static limit appear along the $\gamma$ constants. One can note that the extended Bargmann and the extended para-Bargmann are isomorphic by interchanging the spatial spin-connection with the spatial vielbein. Nevertheless, both theories possess distinct field equations which imply that they are physically different. 

It is interesting to notice that the non-Lorentzian families of CS gravity actions appearing in Table \ref{Table18} can be obtained directly from the relativistic CS action \eqref{Iads}. This can be achieved by expressing the non-Lorentzian gauge fields in terms of the relativistic ones through the semigroup elements as in Table \ref{Table19}. In the case of the extended Para-Bargmann and its flat limit, the corresponding gauge field one-forms can be either obtained from the ultra-relativistic (UR) ones \eqref{ultA} or the non-relativistic (NR) ones \eqref{extA}. Moreover, it is important to mention that the expansion relations of Table \ref{Table19} are not modified in the vanishing cosmological constant limit $\ell\rightarrow\infty$. Thus, the cube in Figure \ref{fig2} is not only valid to relate different kinematical Lie algebras equipped with non-degenerate invariant tensor but also applies to relate the corresponding kinematical CS gravity theories through the $S$-expansion method.
\begin{table}[h!]
\renewcommand{\arraystretch}{1.4}
    \centering
    \begin{tabular}{|l||c|c|c|c|}
    \hline
      \rowcolor[gray]{0.9} Gauge  & Para-Poincaré & Extended  & Extended & Extended
      \\
      \rowcolor[gray]{0.9} fields &  & Newton-Hooke  & para-Bargmann & para-Bargmann
      \\
       \rowcolor[gray]{0.9}  &  &   & (from UR) & (from NR)
      \\
      \hline
        $\omega$ & $\lambda_0 W$ & $\lambda_0 W$ & $\lambda_0 \omega$ & $\lambda_0 \omega$ \\
       \rowcolor[gray]{0.9} $\tau$ & $\lambda_1 V$ & $\lambda_0 V$ & $\lambda_0 \tau$ & $\lambda_1 \tau$ \\
        $\omega_a$ & $\lambda_1 W_a$ & $\lambda_1 W_a$ & $\lambda_1 \omega_a$ & $\lambda_1 \omega_a$ \\
       \rowcolor[gray]{0.9} $e_{a}$ & $\lambda_0 V_a$ & $\lambda_1 V_a$  & $\lambda_1 e_{a}$ & $\lambda_0 e_{a}$ \\
       $s$ & - & $\lambda_2 W$  & $\lambda_2 \omega$  & $\lambda_0 s$\\
        \rowcolor[gray]{0.9} $m$ & - & $\lambda_2 V$  & $\lambda_2 \tau$ & $\lambda_1 m$ \\
     \hline
         \end{tabular}
         \captionsetup{font=footnotesize}
    \caption{Expanded gauge fields in terms of the original ones through the semigroup elements. The multiplication law of the semigroup elements for $S_{E}^{\left(1\right)}$ and $S_{E}^{\left(2\right)}$ are given by \eqref{mlSE1} and \eqref{mlSE2}, respectively.}
    \label{Table19}
\end{table}

As an ending remark, the kinematical CS gravity actions obtained here are based on symmetry algebras equipped with a non-degenerate invariant bilinear form. The non-degeneracy implies that the CS actions ensure a kinetic term for each gauge field and the field equations are given by the vanishing of the curvatures defined in \eqref{relF} and \eqref{extF}. Then, no additional constraint have to be imposed to solve the field equations. The non-degeneracy is not exclusive of the extended kinematical Lie algebras obtained with $S_{E}^{\left(2\right)}$ but appears in every generic cube of Figure \ref{fig4} in which the $S_{E}^{\left(2M\right)}$ semigroup is considered along the speed-space direction. Such particularity can be seen from the non-degeneracy conditions on the relativistic constants $\mu_0$ and $\mu_1$ appearing in the invariant tensor \eqref{IT}. In particular, the non-degeneracy requires $\mu_0\neq\mu_1$ for the AdS algebra and $\mu_1\neq0$ for the Poincaré case. In the non-relativistic regime the $S_{E}^{\left(N\right)}$-expansion of the AdS algebra,  reproducing the $\mathfrak{nh}^{\left(N\right)}$ algebra, admits the following non-vanishing components of the invariant tensor:
\begin{align}
    \langle J^{(n)} J^{(m)} \rangle &= -\alpha_0^{(N)}\,\delta_{N}^{2n+2m} \,, & \langle G_a^{(n)} G_b^{(m)} \rangle &=\alpha_0^{(N)}\,\delta_{N}^{2n+2m+2}\delta_{ab}\,, \notag \\
    \langle J^{(n)}  H^{(m)} \rangle &= -\alpha_1^{(N)}\,\delta_{N}^{2n+2m}\,, & \langle G_a^{(n)} P_b^{(m)}\rangle &=\alpha_1^{(N)}\,\delta_{N}^{2n+2m+2}\delta_{ab}\,,
      \notag \\
    \langle H^{(n)} H^{(m)} \rangle&=-\alpha_0^{(N)}\,\delta_{N}^{2n+2m}\,,  &\langle P_a^{(n)} P_b^{(m)} \rangle&=\alpha_0^{(N)}\,\delta_{N}^{2n+2m+2}\delta_{ab}\,, \label{IT2}
\end{align}
where the $\mathfrak{nh}^{\left(N\right)}$ generators are related to the AdS ones as in Table \ref{Table13}. In the $\mathfrak{gal}^{\left(N\right)}$ limit we have $\langle H^{(n)} H^{(m)} \rangle =\langle P_a^{(n)} P_b^{(m)} \rangle=0$. On the other hand, the constants satisfy
\begin{align}
    \alpha_0^{(N)}&=\lambda_{N}\,\mu_0\,, \\ \alpha_1^{(N)}&=\lambda_{N}\,\mu_1\,,
\end{align}
with $\lambda_N$ being a semigroup element of $S_E^{\left(N\right)}$.
Then, the non-degeneracy of the invariant tensor for the $\mathfrak{nh}^{\left(N\right)}$ algebra implies:
\begin{align}
    \alpha_0^{\left(N\right)}\neq\alpha_1^{\left(N\right)}\,,
\end{align}
where the invariant tensor \eqref{IT2} restricts $N$ to be even. In the $\mathfrak{gal}^{\left(N\right)}$ case, the non-degeneracy conditions reduced to $\alpha_1^{\left(N\right)}\neq 0$ which, from \eqref{IT2}, is satisfied only for even value of $N$. The same analysis applies for the $\mathfrak{pg}^{\left(N\right)}$ and $\mathfrak{stat}^{\left(N\right)}$ obtained as $S_E^{\left(N\right)}$ expansions of generalized Carrollian Lie algebras. The expansion in the speed-space direction restricts the constants appearing in the invariant tensor to exist only for even values of $N$. 

Thus, the next non-degenerate kinematical gravity theory appears for $M=2$, which reproduces the known extended Newtonian gravity \cite{Ozdemir:2019orp}, its Newton-Hooke version \cite{Concha:2019dqs,Concha:2021jos} and its Carrollian counterpart \cite{Gomis:2019nih}. Indeed, as we have discussed in section \ref{sec3}, the $S_E^{\left(3\right)}$-expansion of the Poincaré algebra in the non-relativistic direction reproduces the so-called Newtonian algebra \cite{Hansen:2019pkl}. However, as it was noticed in \cite{Ozdemir:2019orp}, a three-dimensional CS action based on Newtonian symmetry suffers from degeneracy. Such degeneracy has been avoided by introducing an extended version of the Newtonian algebra \cite{Ozdemir:2019orp} which can be obtained as a resonant $S_{E}^{\left(4\right)}$-expansion of the Poincaré algebra along the speed-space direction. The corresponding non-degenerate CS actions based on Post-Newtonian extensions together with their ultra-relativistic versions can be constructed following the same procedure presented in this section.

%%%%%%%%%%%%%%%%%%%%%%%%%%%%%%%%%%%%%%%%%%%%%%%%%%%%%%%%%%%%%%%%%%%%%%%
%%%%%%%%%%%%%%%%%%%%%%%%%%%%%%%%%%%%%%%%%%%%%%%%%%%%%%%%%%%%%%%%%%%%%%%%%%%%

\section{Discussion}\label{sec5}

In this work we have extended the kinematical Lie algebras presented by Bacry and Lévy-Leblond \cite{Bacry:1968zf} through the $S$-expansion procedure. It is well known that the non-relativistic regime of the kinematical algebras presents degeneracy in three spacetime dimensions. Here we have overcome such difficulty by considering expansions instead of contractions to obtain higher-dimensional kinematical algebras which are characterized by additional central charges. Such central charges are crucial to avoid degeneracy and allow to define non-degenerate invariant bilinear trace ensuring the proper construction of a well-defined CS action. We have first showed that the $S_{E}^{\left(1\right)}$ semigroup reproduces the original cube of \cite{Bacry:1968zf}. Then, a non-degenerate cube (see the cube of Figure \ref{fig2}) is obtained in which non-relativistic symmetry algebras are derived by applying a resonant  $S_{E}^{\left(2\right)}$-expansion along the speed-space direction. Subsequently, we have presented generalizations of the kinematical algebras by considering larger semigroups $S_{E}^{\left(N\right)}$ in both speed-time and speed-space directions. Remarkably, the generalized kinematical algebra can be obtained as sequential $S$-expansions starting from $\mathfrak{so}\left(2,2\right)$. 

In the second part of this paper, we have presented the three-dimensional CS gravity actions based on the extended kinematical algebras. The $S$-expansion results are particularly useful to construct the respective CS actions since they provide us with the non-vanishing components of the expanded invariant tensor in terms of the original ones. Moreover, the extended kinematical algebras admit a non-degenerate invariant tensor leading to field equations that are given by the vanishing of the curvature two-forms for the extended kinematical algebras in the cube of Figure \ref{fig2}. Then, no additional constraints have to be imposed to solve the equations of motion. Finally, we showed that the non-Lorentzian gauge fields can be expressed in terms of the relativistic ones through the semigroup elements, allowing us to obtain each kinematical CS action from the relativistic one using the semigroup expansion properties.

The generalization of our results to supersymmetry is a natural next step. In particular, the expansion procedure based on semigroups would allow us not only to obtain a supersymmetric extension of the original cube of Bacry and Lévy-Leblond \cite{Bacry:1968zf} but also to obtain the extended kinematical superalgebra which admits non-degenerate invariant bilinear trace. As it was shown in \cite{Bergshoeff:2016lwr,Ozdemir:2019tby,Concha:2020eam}, in addition to the extra bosonic content, supplementary fermionic charges have to be added to derive the supersymmetric extension of the extended Bargmann symmetry algebra in the absence of degeneracy. Our method could provide us a general recipe to obtain non-degenerate kinematical supergravity theories as subsequential expansions starting from the AdS supergravity. On the other hand, the number and nature of the extra gauge fields appearing in the supersymmetric extensions could offer a criteria to select relativistic theories. 

Another aspect that deserve to be explored is the study of the kinematical algebras considering symmetry algebras beyond AdS and its Poincaré limit. A natural deformation of Poincaré algebra is the Maxwell algebra \cite{Bacry:1970du,Bacry:1970ye,Schrader:1972zd,Gomis:2017cmt} whose physical implications in gravity have been extensively studied \cite{Cangemi:1992ri,Duval:2008tr,Gomis:2009dm,Bonanos:2010fw,deAzcarraga:2010sw,Durka:2011nf,deAzcarraga:2012qj,Salgado:2014jka,Hoseinzadeh:2014bla,Concha:2014tca,Ravera:2018vra,Concha:2018zeb,Concha:2018jxx,Concha:2019icz,Salgado-Rebolledo:2019kft,Chernyavsky:2020fqs,Adami:2020xkm,Kibaroglu:2020tbr,Cebecioglu:2022iyq,Durka:2022ghv,Kibaroglu:2022bzq,Caroca:2023oie}. Unlike the Poincaré algebra, the Maxwell algebra is benefited with a non-degenerate invariant metric even in higher spacetime dimensions \cite{Soroka:2004fj,Matulich:2019cdo}. In this direction one could obtain deformed kinematical algebras admitting non-degenerate invariant tensor in arbitrary spacetime-dimensions. In three spacetime dimensions, one expect to obtain not only known non-Lorentzian Maxwellian symmetry algebras \cite{Aviles:2018jzw,Gomis:2019nih,Concha:2020sjt,Concha:2020ebl,Concha:2021jnn} but also new ones. 

It would also be interesting to explore which kinematical algebras can admit a non-vanishing torsion. The inclusion of a torsion in the relativistic theory can be done through the Mielke-Baekler gravity formalism \cite{Mielke:1991nn,Baekler:1992ab} which implies the modification of the Riemannian geometry to a Weizenböck one. At the non-relativistic level, torsional gravity models have been recently proposed in \cite{Concha:2021llq,Concha:2022you,Concha:2023ejs} by considering deformations of the extended Bargmann and extended Newtonian symmetry algebras. One could analyze if the torsion can be switched on for other (generalized) kinematical algebras and study the relation with the torsional Newton-Cartan gravity model introduced in \cite{Bergshoeff:2015ija,Bergshoeff:2017dqq,VandenBleeken:2017rij}.

It would be worth it to explore if our results can be generalized to higher-spin algebras. The extensions of the original kinematical algebras \cite{Bacry:1968zf} to three-dimensional theories that include a spin-3 gauge field coupled to gravity have already been presented in \cite{Bergshoeff:2016soe,Concha:2022muu,Caroca:2022byi}. Our results could allows us not only to reproduce known results but also to obtain the minimal non-degenerate cube required to construct well-defined spin-3 CS gravity actions. Then, one could apply our procedure to explore the kinematical algebras for theory containing massless spin-$\frac{5}{2}$ gauge field. Thus, higher-spin kinematical algebra could be obtained as sequential expansions starting from the $\mathfrak{osp}\left(1|4\right)\times\mathfrak{osp}\left(1|4\right)$ and the hyper-Poincaré algebra. Moreover, the $S$-expansion method could provides us with the desired general kinematical algebras equipped with non-degenerate invariant tensor allowing us to construct proper CS hypergravity theory.

\section*{Acknowledgments}

This work was funded by the National Agency for Research and Development ANID - SIA grant No. SA77210097 and FONDECYT grants No. 1211077, 11220328 and 11220486. This work was supported by USC20102 Internacionalización Transversal en la UCSC: enfrentando los nuevos desafíos (P.C.).  D.P. would like to thank to Universidad de Concepción, Chile, for Beca articulación pregrado-postgrado. L.R. would like to thank the DISAT of the Polytechnic of Turin and the INFN for financial support. P.C., L.R. and E.R. would like to thank to the Dirección de Investigación and Vice-rectoría de Investigación of the Universidad Católica de la Santísima Concepción, Chile, for their constant support.

%%%%%%%%%%%%%%%%%%%%%%%%%%%%%%%%%%%%%%%%%%%%%%%%%%%%%%%%%%%%%%%%
%%%%%%%%%%%%%%%%%%%%%%%%%%%%%%%%%%%%%%%%%%%%%%%%%%%%%%%%%%%%%%%%%%%%%%%%%%%%

\bibliographystyle{fullsort.bst}
 
\bibliography{Non_Lorentzian_expansion}

\end{document}